\documentclass[lettersize,journal]{IEEEtran}
\usepackage{amsmath,amsfonts}
\usepackage{algorithmic}
\usepackage{algorithm}
\usepackage{array}
\usepackage[caption=false,font=normalsize,labelfont=sf,textfont=sf]{subfig}
\usepackage{textcomp}
\usepackage{stfloats}
\usepackage{multirow}
\usepackage{url}
\usepackage{verbatim}
\usepackage{graphicx}
\usepackage{cite}
\usepackage{acronym}
\usepackage{threeparttable}
\begin{document}

\begin{acronym}
        \acro{DNNs}{Deep Neural Networks}
        \acro{WNNs}{Weightless Neural Networks}
        \acro{DL}{Deep Learning}
        \acro{CPUs}{Central Processing Units}
        \acro{GPUs}{Graphics Processing Units}
        \acro{FPGAs}{Field-Programmable Gate Arrays}
        \acro{LUT}{Lookup Table}
        \acro{LUTs}{Lookup Tables}
        \acro{BNNs}{Binary Neural Networks}
        \acro{P-LUTs}{Physical-LUTs}
        \acro{L-LUTs}{Logical-LUTs}
        \acro{MLPs}{Multilayer Perceptrons}
        \acro{GBDTs}{Gradient Boosted Decision Trees}
        \acro{EFD}{Extended Finite Difference}
        \acro{VC}{Vapnik-Chervonenkis}
        \acro{DSPs}{Digital Signal Processings}
        \acro{DSP}{Digital Signal Processing}
        \acro{MACs}{Multiply-Accumulates}
        \acro{ISFs}{Incompletely Specified Functions}
        \acro{LLM}{Large Language Model}
        \acro{CIM}{Computing In-Memory}
        \acro{ReLU}{Rectified Linear Units}
\end{acronym}

\title{A Survey on LUT-based Deep Neural Networks Implemented in FPGAs}

\author{\IEEEauthorblockN{1\textsuperscript{st} Zeyu Guo}
\IEEEauthorblockA{
\textit{Huazhong University of Science and Technology\\
Wuhan, China \\
zanek8zc@outlook.com}
\thanks{I would like to express my sincere gratitude to \textbf{sigma\_delta66} for his patient guidance and insightful suggestions, which greatly improved the quality of this work.}
}}



\maketitle
\begin{abstract}

Low-latency, energy-efficient deep neural networks (DNNs) inference are critical for edge applications, where traditional cloud-based deployment suffers from high latency and security risks. Field-Programmable Gate Arrays (FPGAs) offer a compelling solution, balancing reconfigurability, power efficiency, and real-time performance. However, conventional FPGA-based DNNs rely heavily on digital signal processing (DSP) blocks for multiply-accumulate (MAC) operations, limiting scalability.

LUT-based DNNs address this challenge by fully leveraging FPGA lookup tables (LUTs) for computation, improving resource utilization and reducing inference latency. This survey provides a comprehensive review of LUT-based DNN architectures, including their evolution, design methodologies, and performance trade-offs, while outlining promising directions for future research.
\end{abstract}


\section{Introduction}

\ac{DNNs} have demonstrated significant capabilities in feature extraction and regression, making them widely applicable across various domains, for example, in particle collision\cite{duarteFastInferenceDeep2018a}, image recognition, and natural language processing\cite{shuvoEfficientAccelerationDeep2023}. However, deploying these models in cloud-based environments often results in challenges such as the inability to achieve real-time performance, high security risks, and increased latency. Consequently, there has been a growing interest in deploying DNNs on edge devices. Nevertheless, the high computational complexity, substantial power consumption, and extensive memory requirements of these models pose significant challenges for edge deployment.

In the context of edge computing, \ac{CPUs}, \ac{GPUs}, and \ac{FPGAs} represent the three primary hardware platforms for DNN deployment, each with distinct advantages and limitations. CPUs, while highly flexible and capable of handling diverse workloads, generally lack the parallelism required for efficient DNN inference, resulting in suboptimal performance for real-time applications. GPUs, leveraging thousands of parallel cores, significantly accelerate matrix operations and have become the standard choice for DNNs. However, their high power consumption, memory bandwidth limitations, and relatively large physical footprint make them less ideal for edge deployment, where energy efficiency and latency are critical constraints. FPGAs, by contrast, offer a unique balance between performance, power efficiency, and flexibility. Their reconfigurable nature allows for hardware customization tailored to specific DNN architectures, enabling low-latency and energy-efficient inference. Moreover, FPGAs can be optimized for specialized dataflows, achieving high throughput with minimal resource usage, making them an attractive choice for edge-based DNN deployment.

To execute the extensive multiply-accumulate (MAC) operations required in DNNs, the most straightforward approach is utilizing the \ac{DSP} resources within FPGAs. However, FPGA-based DSP resources are highly constrained, and even with techniques such as quantization and pruning, DSP-based implementations still consume a significant portion of available DSP blocks, limiting the scalability of such designs. To reduce DSP utilization in MAC operations, Binarized Neural Networks (BNNs) were introduced, which replace multiplications with XNOR operations by binarizing both input vectors and weight parameters. Despite reducing DSP usage, XNOR operations mapped to FPGA hardware primarily utilize 2-input \ac{LUTs}, which are treated as simple logic gates, failing to fully exploit the computational potential of FPGA architectures.

Given that DSP-based DNNs and XNOR-based BNNs do not fully harness the computational advantages of FPGAs, exploring LUT-based implementations for higher-performance DNNs is a meaningful research direction. For instance, approaches such as LogicNets\cite{umurogluLogicNetsCoDesignedNeural2020} and PolyLUT\cite{andronicPolyLUTLearningPiecewise2023b} encapsulate all operations between quantized inputs and quantized outputs within LUTs, achieving higher resource utilization and lower latency. However, a comprehensive survey summarizing the development and technical advancements of LUT-based DNNs is still lacking. To address this gap, this survey provides a detailed review of LUT-based DNNs, tracing their evolution and technological pathways while conducting experimental comparisons of their performance. Finally, potential future research directions are outlined to further advance this field.

\section{Conventional Implementations of Deep Neural Networks in FPGAs}

\subsection{DSP-based Deep Neural Networks}

\begin{figure}[t]
\centering
\includegraphics[width=3.5in]{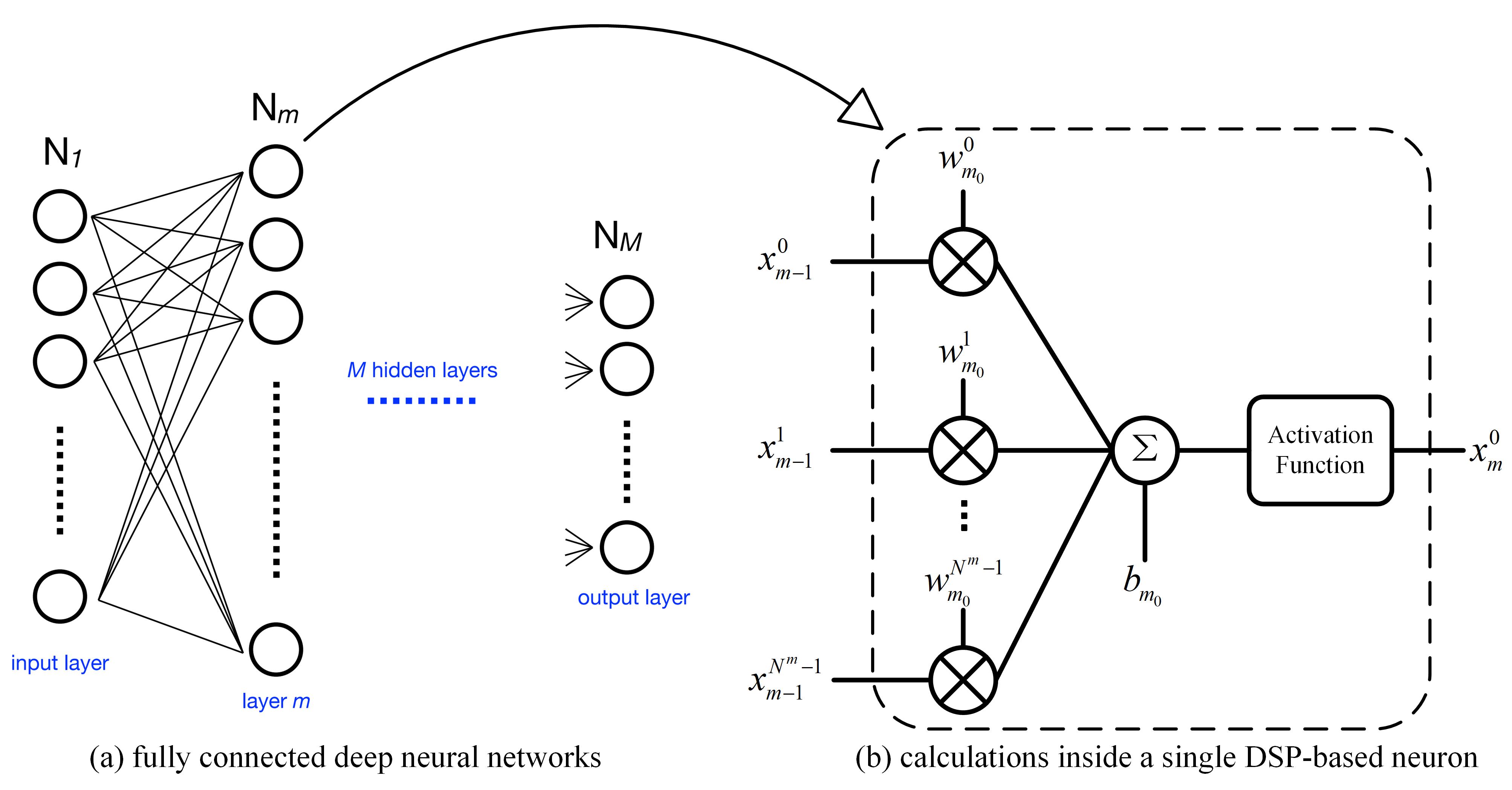}
\caption{Architectures of DSP-based fully connected DNNs and a single neuron\cite{duarteFastInferenceDeep2018a}.}
\label{dsp}
\end{figure}

DNNs are network structures composed of multiple layers of interconnected artificial neurons, typically consisting of an input layer, one or multiple hidden layers, and an output layer. As shown in Figure \ref{dsp}(a), a fully connected DNN consists of \( M \) layers, where the input layer contains \( N_1 \) input neurons, each of the \( m \) hidden layers comprises \( N_m \) neurons, and the output layer consists of \( N_M \) output neurons. The internal function of each neuron is generally composed of a series of linear transformations followed by an activation function. For instance, the output \( x_m^0 \) of the first neuron in the \( m \)-th layer can be expressed as:  

\begin{equation}
\label{full-con}
x_{m}^{0} = \sigma \left[ \sum_{i = 0}^{N_{m-1}-1} w_{m_{0}}^{i}x_{m-1}^{i} + b_{m_{0}} \right],
\end{equation}

where \( x_{m-1} \) represents the output feature vector of layer \( m-1 \), \( w_{m_{0}} \) denotes the weight parameter vector of the first neuron in layer \( m \), \( b_{m_{0}} \) is the bias parameter, and \( \sigma \) is the activation function (e.g., \ac{ReLU} or Softmax). The most straightforward approach of performing the linear transformation in the above equation is through MAC operations, as illustrated in Figure \ref{dsp}(b). DSP-based DNN implementations leverage the DSP resources in FPGAs to perform MAC computations.  

\texttt{hls4ml} \cite{duarteFastInferenceDeep2018a} is an open-source architecture designed for deploying low-latency DNNs on FPGAs. Duarte \textit{et al.} \cite{duarteFastInferenceDeep2018a} utilized the \texttt{hls4ml} framework, integrating quantization, pruning, and parallelization techniques to achieve low-latency DNN designs with both fully unrolled and partially unrolled implementations. Quantization refers to converting weight parameters and activation outputs from floating-point precision (e.g., float32) to integer precision (e.g., int8), significantly reducing resource consumption and inference latency. Pruning is a post-training technique that eliminates low-weight connections, increasing the sparsity of neuron inputs, thereby further reducing hardware resource consumption.  

Despite quantization and pruning, DSP-based DNNs still consume a significant number of DSP resources. However, the number of DSP units available within an FPGA is highly limited, restricting the capability of DSP-based DNN implementations to handle more complex tasks. This constraint has motivated researchers to explore alternative methods for reducing the resource consumption of FPGA-based DNN implementations.  

\subsection{XNOR-based Deep Neural Networks}

\begin{figure}[h]
\centering
\includegraphics[width=1.5in]{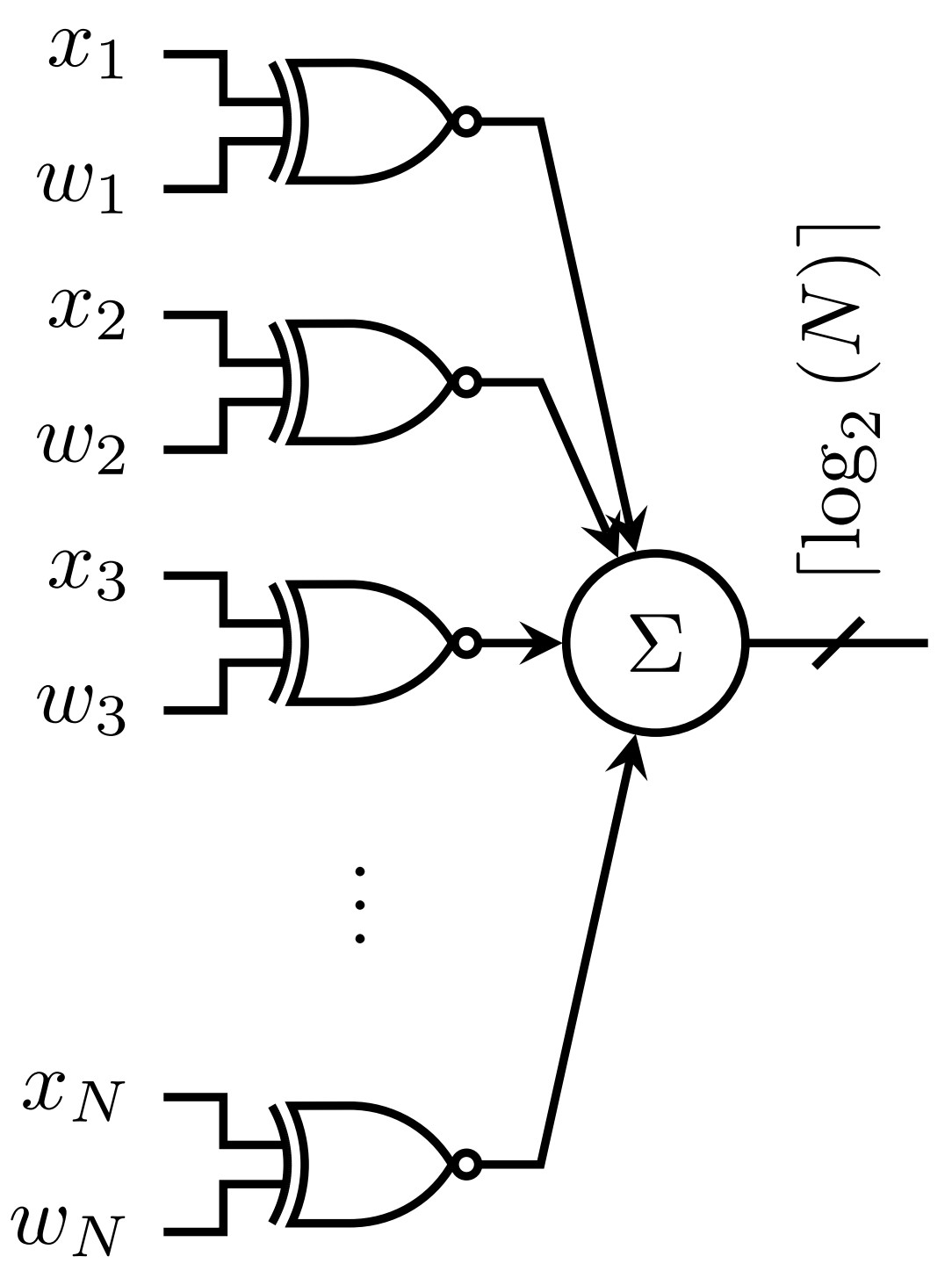}
\caption{Architecture of a XNOR-based neuron\cite{wangLUTNetRethinkingInference2019}.}
\label{bnn}
\end{figure}

To reduce the DSP resources consumed by MAC operations, researchers have proposed the BNN, which binarizes the weight parameters (\( w \in \{-1,1\} \)). As illustrated in Figure \ref{bnn}, BNN neuron architectures simplify MAC operations—multipliers can be replaced with XNOR gates, while adders can be replaced with population count (popcount) operations (which count the number of ones in the XNOR results) \cite{umurogluFINNFrameworkFast2017}. In BNNs, XNOR operations are implemented using 2-input LUTs in FPGA hardware. This significantly reduces memory and computation demands, making model deployment on resource-constrained devices much more feasible. Given an input vector \( \mathbf{x} = [x_1, x_2, \dots, x_N] \), a binarized weight vector \( \mathbf{w} = [w_1, w_2, \dots, w_N] \), and an activation function \( \sigma \), the output can be expressed as:  

\begin{equation}
\label{bnn-eq1}
y = \sigma \left( \sum_{i=1}^{N} \text{XNOR}(x_i, w_i) \right)
\end{equation}

\begin{equation}
\label{bnn-eq2}
\text{XNOR}(x_i, w_i) = \begin{cases}
        0, & \text{if } x_i \neq w_i, \\
        1, & \text{if } x_i = w_i.
        \end{cases} 
\end{equation}

Even though popcount operations reduce computational complexity, the results of all XNOR operations still need to be accumulated to obtain the final output. In modern DNNs, the number of input features \( N \) often reaches hundred even thousand, leading to significant resource consumption. Moreover, XNOR-based computations are limited to binary multiplications and cannot fully utilize the LUT resources of FPGAs, as they treat LUTs merely as simple logic gates.

\section{LUT-based Deep Neural Networks in FPGAs}

\subsection{LUTNet}

\begin{figure}[t]
\centering
\includegraphics[width=2.5in]{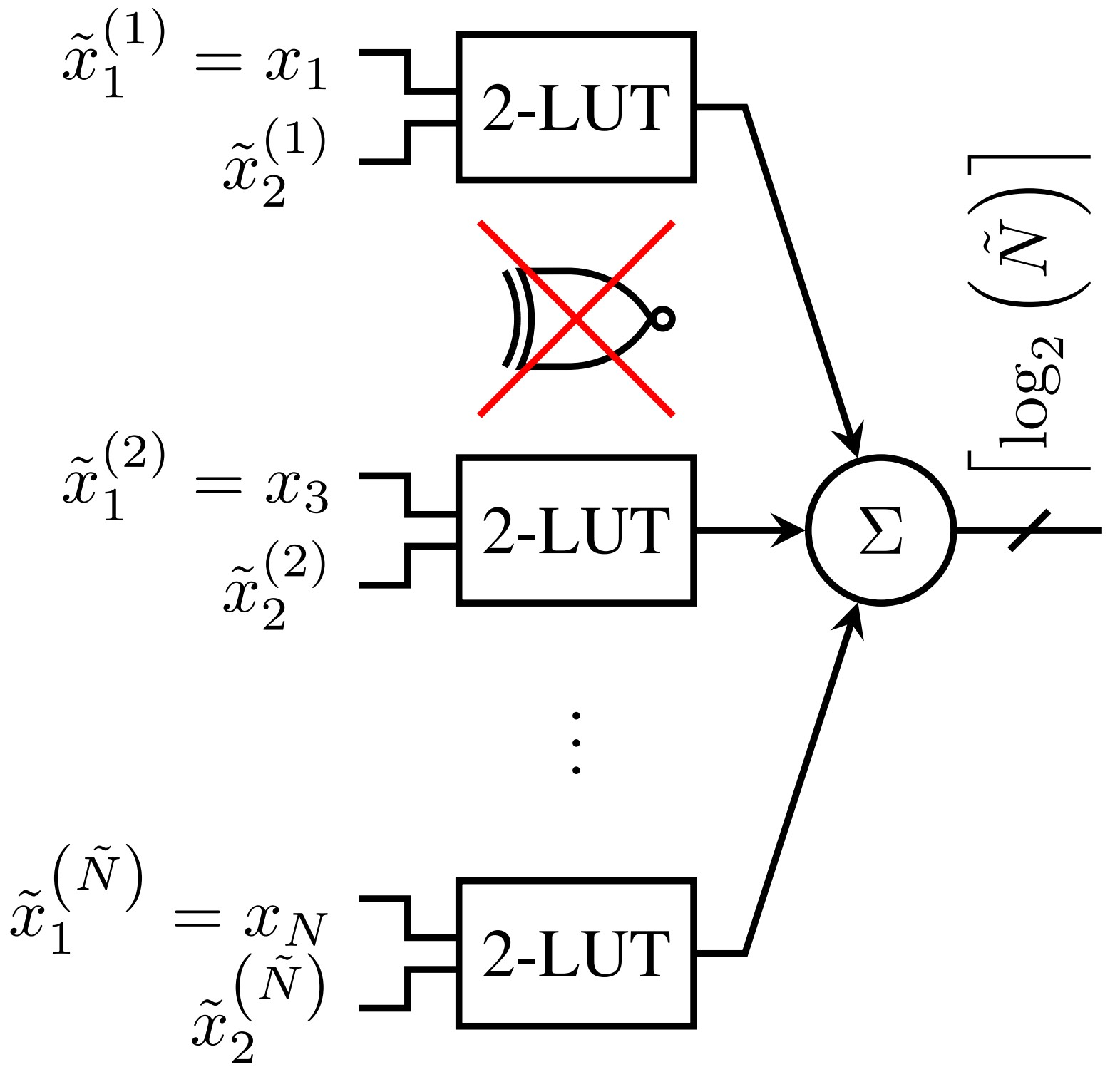}
\caption{Architecture of a LUTNet neuron\cite{wangLUTNetRethinkingInference2019}.}
\label{lutnet}
\end{figure}

To address the limitations of XNOR-based DNNs, LUTNet \cite{wangLUTNetRethinkingInference2019} proposes replacing XNOR computations with LUT-based implementations. Figure \ref{lutnet} illustrates the LUTNet neuron architecture. The core idea of LUTNet is to leverage the \( K \)-input lookup tables (K-LUTs) in FPGAs to implement arbitrary \( K \)-input Boolean functions. Instead of traditional XNOR operations, LUTNet trains specific \( K \)-LUT logic functions \( g_n \) to replace them:

\begin{equation}
\label{lutnet-eq}
y = \sigma \left( \sum_{n=1}^{\widetilde{N}} g_n(\widetilde{x}^{(n)}) \right),
\end{equation}
where \( \sigma \) is the activation function, \( \widetilde{N} \) represents the number of pruned connections (before pruning, it was \( N \)), typically \( \widetilde{N} \ll N \). The term \( \widetilde{x}^{(n)} \) represents a randomly selected subset of \( K \) elements from the input vector, serving as the input to the \( K \)-LUT function \( g_n \). Each trained \( g_n \) is directly mapped to FPGA \( K \)-LUTs, rather than simply executing XNOR operations.  BNNs can be considered a special case of LUTNet when \( K = 1 \) and \( \widetilde{N} = N \).  

Compared to BNNs, LUTNet offers several advantages. First, during training, \( K \)-input LUTs can learn more complex nonlinear relationships instead of being limited to individual XNOR computations. Additionally, for large input sets, LUTNet can further reduce adder resource consumption. Traditional BNNs require summing the results of \( N \) XNOR operations, whereas LUTNet groups computations using \( K \)-LUTs, reducing the need for a global popcount operation. For example, when \( N = 12 \), a conventional BNN requires 12 two-input LUTs and a large popcount tree, while LUTNet only needs three 4-input LUTs for partial summation, followed by a smaller popcount operation. This significantly reduces adder usage. Lastly, LUTNet improves FPGA resource efficiency by pruning unimportant computation paths, thereby reducing the number of LUT inputs while maintaining the same inference accuracy.  

Despite its advantages, LUTNet still relies on popcount operations, which in turn leads to substantial resource consumption. Moreover, the linear activation function lacks non-linearity, limiting a DNN's ability to learn complex patterns and extract hierarchical features.

\subsection{NullaNet}

\begin{figure}[t]
\centering
\includegraphics[width=3.5in]{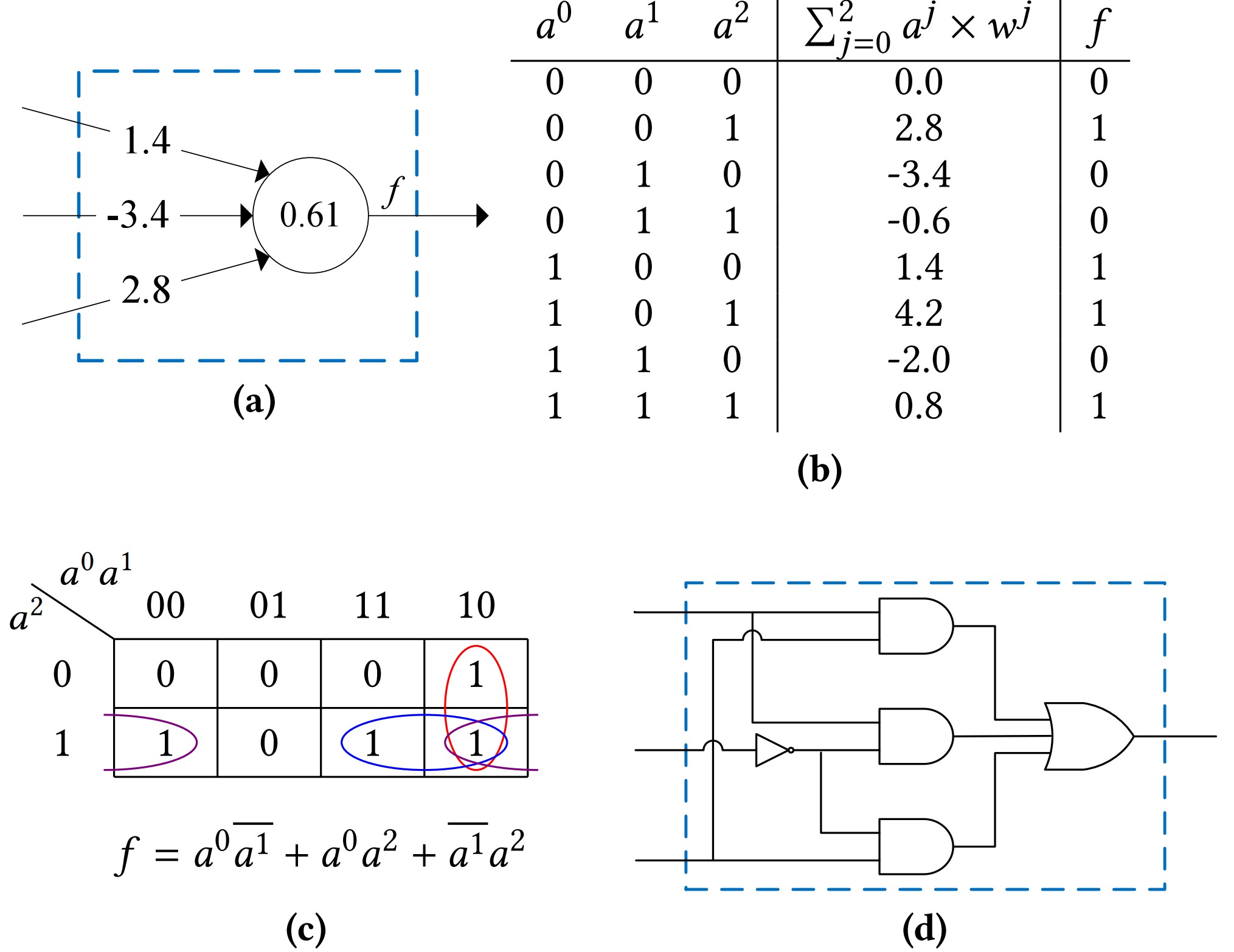}
\caption{(a) Architecture of a NullaNet neuron. (b) Examples of McCulloch-Pitts neuron modeling calculations. (c) Karnaugh map simplification. (d) Implementation using logic gates after Karnaugh map simplification.\cite{nazemiEnergyefficientLowlatencyRealization2019}.}
\label{nullalut}
\end{figure}

To further optimize resource utilization, NullaNet \cite{nazemiEnergyefficientLowlatencyRealization2019} introduced the idea of implementing all computations within a neuron using Boolean logic functions, which can be efficiently realized using LUT resources. Figure \ref{nullalut}(a) illustrates the NullaNet neuron architecture, where trained binarized neurons are optimized using Boolean logic functions. For example, each neuron’s output can be modeled by a McCulloch-Pitts neuron:

\begin{equation}
\label{nulla}
f =
\begin{cases}
1, & \sum_j a_j \cdot w_j \geq b \\
0, & \text{otherwise}
\end{cases}
\end{equation}
where \( a_j \) and \( w_j \) are the input and weight values, respectively, and \( b \) is the threshold. The McCulloch-Pitts model results in the computation shown in Figure \ref{nullalut}(b).  

\begin{figure}[t]
\centering
\includegraphics[width=3.5in]{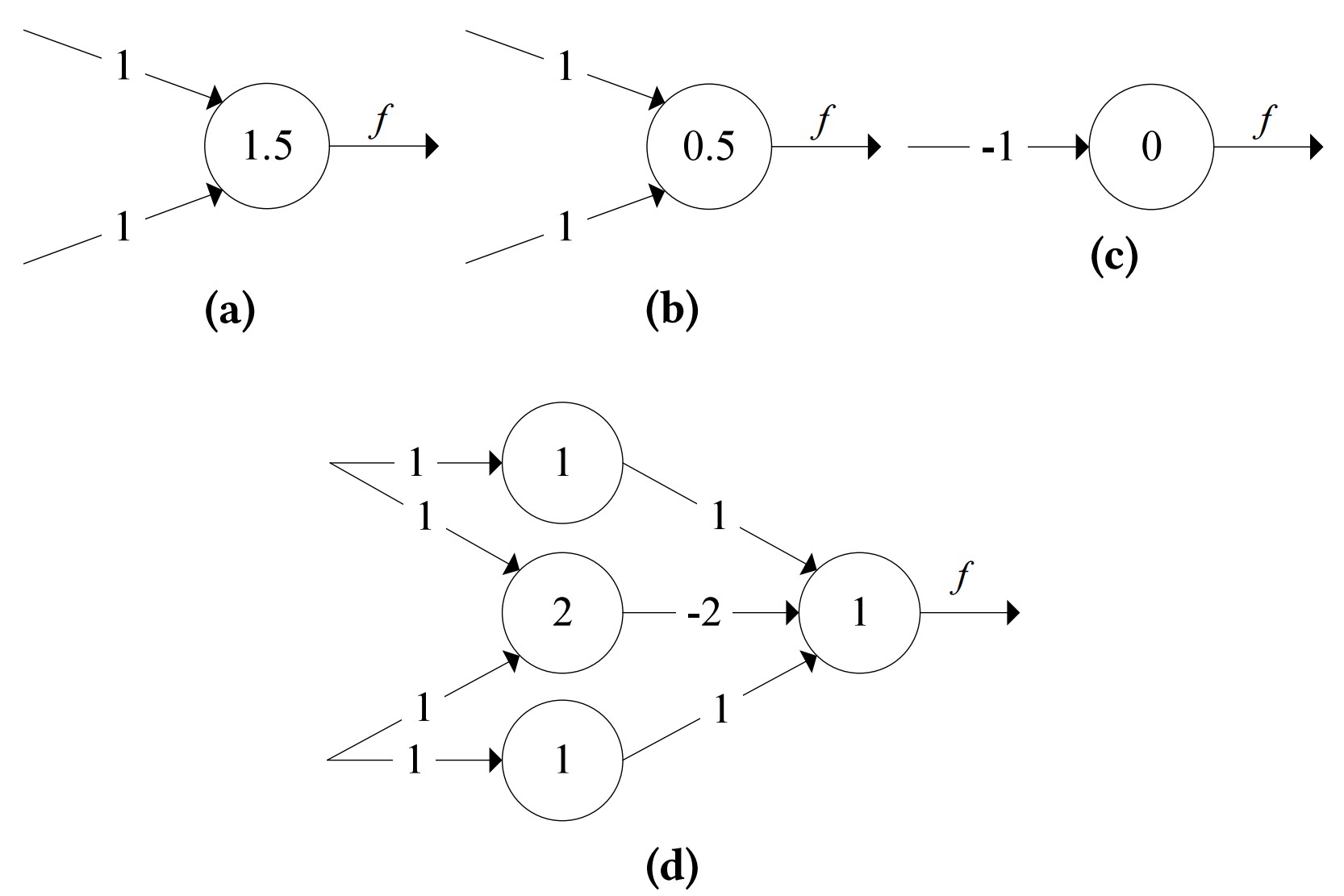}
\caption{Examples of implementing logic gates using McCulloch-Pitts neurons: (a) AND, (b) OR, (c) NOT, (d) XNOR\cite{nazemiEnergyefficientLowlatencyRealization2019}.}
\label{nullalut-2}
\end{figure}

For neurons with a small number of inputs, a truth table optimization approach can be applied. As depicted in Figure \ref{nullalut}(a), the neuron’s logic function can be minimized using Karnaugh maps (K-Maps), leading to an optimized Boolean expression that can be implemented using the logic gates in Figure \ref{nullalut}(d). Figure \ref{nullalut-2} further illustrates that these logic gates can be modeled by McCulloch-Pitts neurons. Consequently, NullaNet transforms traditional neuron computations into Boolean logic operations using logic gates. However, for neurons with a large number of inputs, the size of the truth table grows exponentially, making this approach impractical for DNNs with many neurons.

To mitigate this issue, NullaNet introduces the concept of Incompletely Specified Functions (ISFs) to approximate DNN computations, significantly reducing resource consumption. In Boolean algebra, a Completely Specified Function (CSF) is a Boolean function that has a defined output (0 or 1) for all possible input combinations. In contrast, an ISF allows some input combinations to have unspecified outputs. An ISF consists of the following three sets:  

\begin{itemize}
        \item ON-set: Input combinations for which the function outputs 1.
        \item OFF-set: Input combinations for which the function outputs 0.  
        \item Don't-Care Set (DC-set): Input combinations with undefined outputs, which can be optimized to simplify the circuit structure and reduce logic complexity.
\end{itemize}

Compared to LUTNet, NullaNet achieves higher resource efficiency by storing optimized Boolean logic functions directly in lookup tables, effectively integrating all neuron computations within LUTs.

However, NullaNet has certain limitations. The ISF method is a form of lossy truth table sampling, meaning that for unseen inputs, it cannot strictly guarantee that its output matches the original DNN’s pre-optimized result. Instead, it provides an approximate inference. Furthermore, NullaNet has only been applied to fully connected DNNs. Although ISFs reduce resource consumption, the scalability of the network remains constrained by the number of neuron inputs. Additionally, NullaNet only considers binary quantization, limiting its overall network accuracy.

\subsection{LogicNets}

To address the accuracy loss caused by the lossy truth table sampling in NullaNet, LogicNets\cite{umurogluLogicNetsCoDesignedNeural2020} generates logical lookup tables (L-LUTs) to enumerate all function values arising from the quantized inputs and outputs. The resulting netlist is then mapped onto the physical LUTs (P-LUTs) resources of an FPGA through logic synthesis tools, ensuring no accuracy loss during the mapping process from the quantized theoretical model to the actual hardware.

\begin{figure}[b]
\centering
\includegraphics[width=3in]{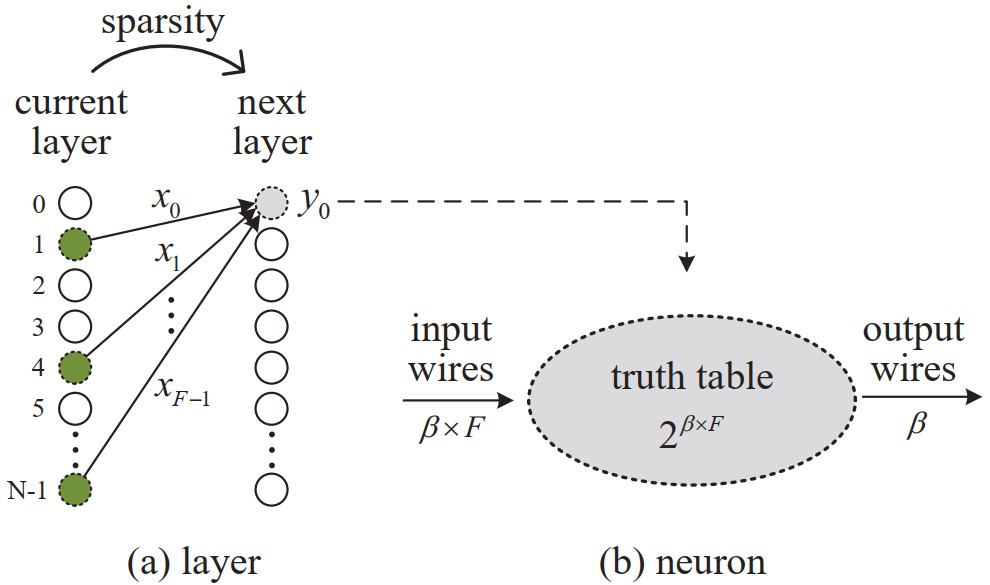}
\caption{High sparsity connection scheme of LogicNet\cite{umurogluLogicNetsCoDesignedNeural2020}.}
\label{sparse}
\end{figure}

While NullaNet considers only fully connected DNNs with binary quantization, LogicNets uses high sparse connectivity to reduce the large fan-in based on expander graph theory. As illustrated in Figure \ref{sparse}, the sparse connectivity method in LogicNets randomly selects only \(F\) neurons \([x_{0}, x_{1}, ... x_{F-1} ]\) out of \(N\) neurons in the current layer to connect to the input of neuron \(y_{0}\) in the next layer. To minimize the size of the lookup table, \(F\) must be significantly smaller than \(N\) (\(F \ll N\)). Additionally, LogicNets is not restricted to binary quantization but support feature vectors with a bit width of \(\beta\). Consequently, the input bit width of the generated lookup tables is \(\beta F\), leading to a table size of \(2^{\beta F}\).

\begin{figure}[t]
\centering
\includegraphics[width=2in]{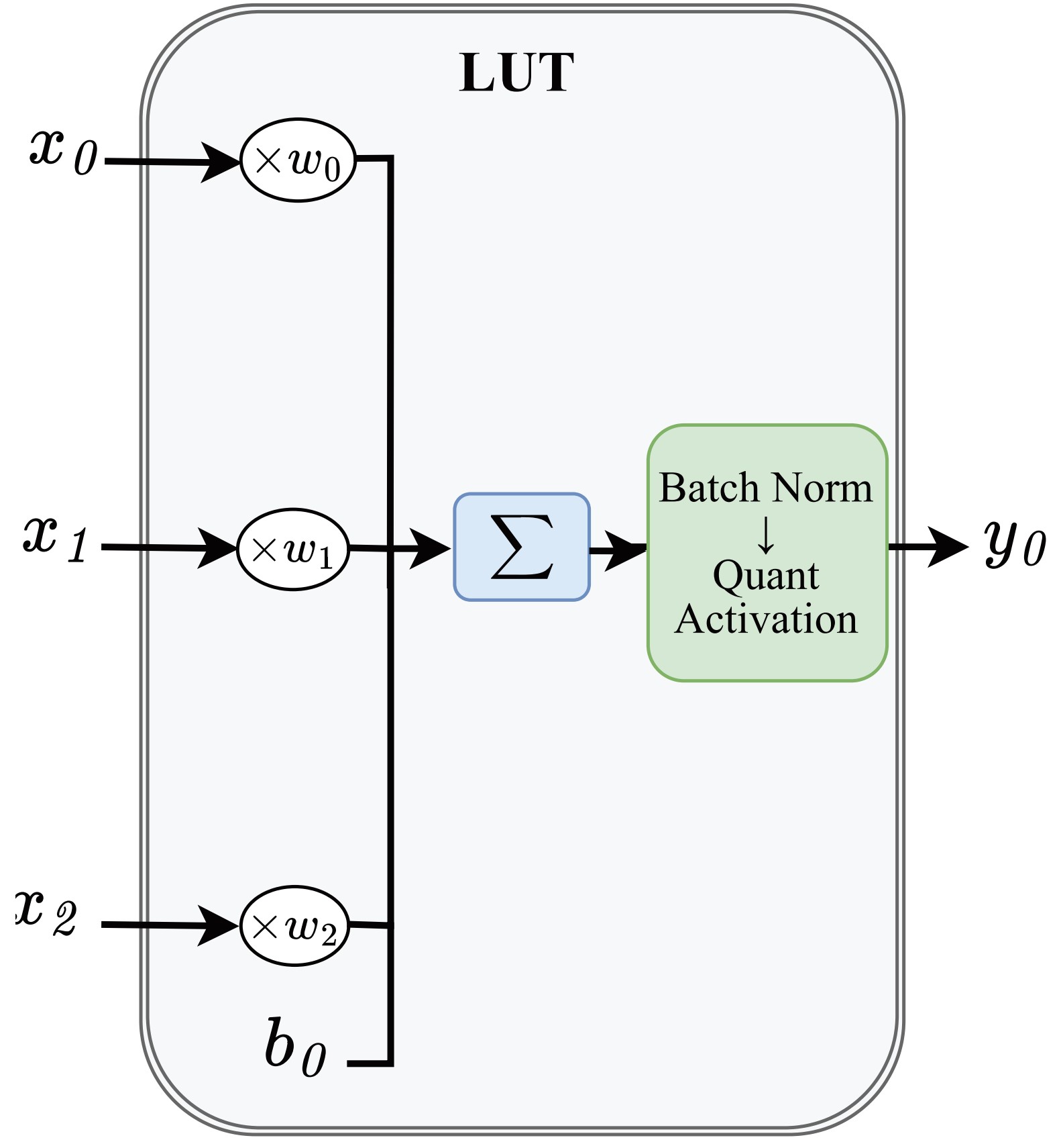}
\caption{Architecture of a LogicNets neuron\cite{andronicPolyLUTLearningPiecewise2023b}.}
\label{logicnets}
\end{figure}

Figure \ref{logicnets} presents the neural architecture of LogicNets when \(F = 3\), where the neuron output \(y_{0}\) can be expressed as:

\begin{equation}
\label{linear}
y_{0} = \sigma \left[ \sum_{i = 0}^{F-1} w_{i}x_{i} + b \right],
\end{equation}
where \(\sigma\) represents the activation function, \(w\) is the weight vector, and \(b\) is the bias parameter.

Compared to NullaNet, LogicNets adopts a lossless approach that encapsulates all operations between quantized inputs and outputs into lookup tables. By integrating high sparse connectivity with low-bit quantization, LogicNets achieves higher accuracy and lower latency in DNN implementation.

Despite these advantages, LogicNets suffers from an exponential increase in lookup table size as the number of neuron inputs grows. Furthermore, the model employs only linear transformations and activation functions for training and inference, limiting its capability in predicting complex tasks.

\subsection{PolyLUT}

\begin{figure}[t]
        \centering
        \includegraphics[width=2.5in]{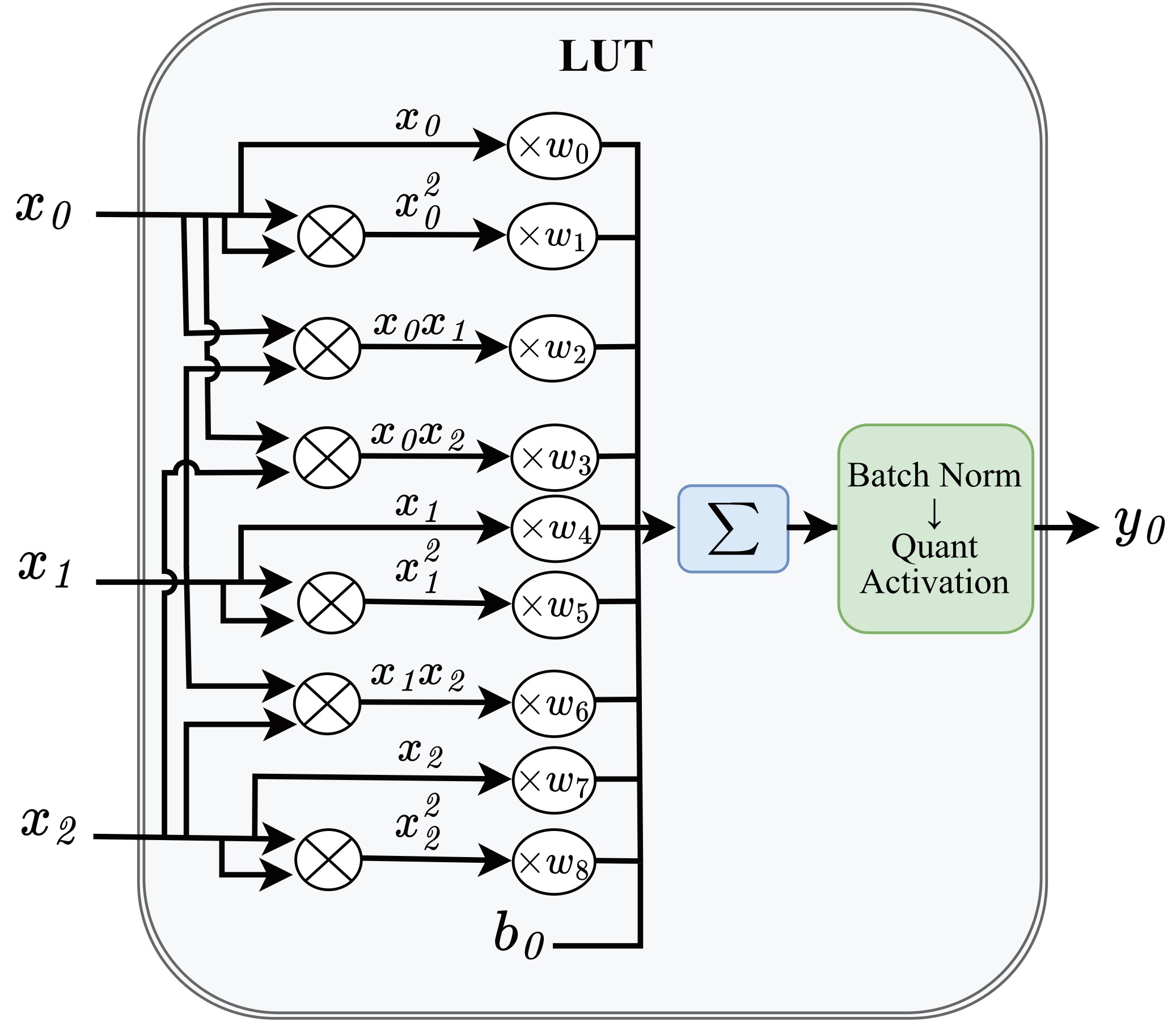}
        \caption{Architecture of a PolyLUT neuron\cite{andronicPolyLUTLearningPiecewise2023b}.}
        \label{polylut}
\end{figure}

DNNs function as universal approximators, meaning that with a sufficient number of neurons (either in depth or width), they can approximate any continuous function. However, LogicNets approximates functions using only linear transformations, restricting the feature vector width to the sparse neuron inputs \(F\), which constrains its capacity for complex task prediction. To overcome this limitation, PolyLUT\cite{andronicPolyLUTLearningPiecewise2023b} replaces the linear transformations in LogicNets with multivariate polynomial functions. By introducing an adjustable polynomial degree \(D\), PolyLUT extends the feature vector width from \(F\) to \(\frac{(F+D)!}{F!D!}\), enhancing function approximation capability. The sparse pruning method from LogicNets is retained for inter-layer connections. For instance, as shown in Figure \ref{polylut}, when \(F = 3\) and \(D = 2\), the input feature vector expands from \([x_{0}, x_{1}, x_{2}]\) to \([1, x_{0}, x_{1}, x_{2}, x_{0}x_{1}, x_{0}x_{2}, x_{1}x_{2}, x_{0}^2, x_{1}^2, x_{2}^2]\). Notably, despite using more complex multivariate polynomial functions, PolyLUT does not increase the consumption of L-LUT resources. Since the size of L-LUTs is determined by the neuron input \(F\) and input bit width \(\beta\), and both remain unchanged, PolyLUT maintains the same L-LUT size as in linear transformation, i.e., \(2^{\beta F}\). The neuron output \(y_{0}\) is given by:

\begin{equation}
\label{polynomial}
y_{0} = \sigma \left[ \sum_{i = 0}^{M-1} w_{i}m_{i}(x) + b \right], \quad \mathrm{where}\;M = \binom{F+D}{D}
\end{equation}
where \(\sigma\) is the activation function, \(w\) is the weight vector, \(b\) is the bias parameter, and \(M\) represents the number of monomials \(m(x)\) of degree at most \(D\) in \(F\) variables.

Unlike traditional DNN hardware accelerators, where replacing linear transformations with polynomial functions increases the demand for multipliers or XNOR gates, LUT-based DNN implementations encapsulate the additional multiplications within LUTs. This enables PolyLUT to achieve comparable inference accuracy with fewer network layers or neurons, resulting in reduced resource consumption and lower latency.

However, PolyLUT still suffers from an exponential increase in lookup table size with growing neuron inputs, limiting its scalability. Additionally, as the polynomial degree \(D\) increases, training becomes more challenging, potentially leading to unstable optimization or overfitting.

\subsection{PolyLUT-Add}

\begin{figure}[t]
        \centering
        \includegraphics[width=3.5in]{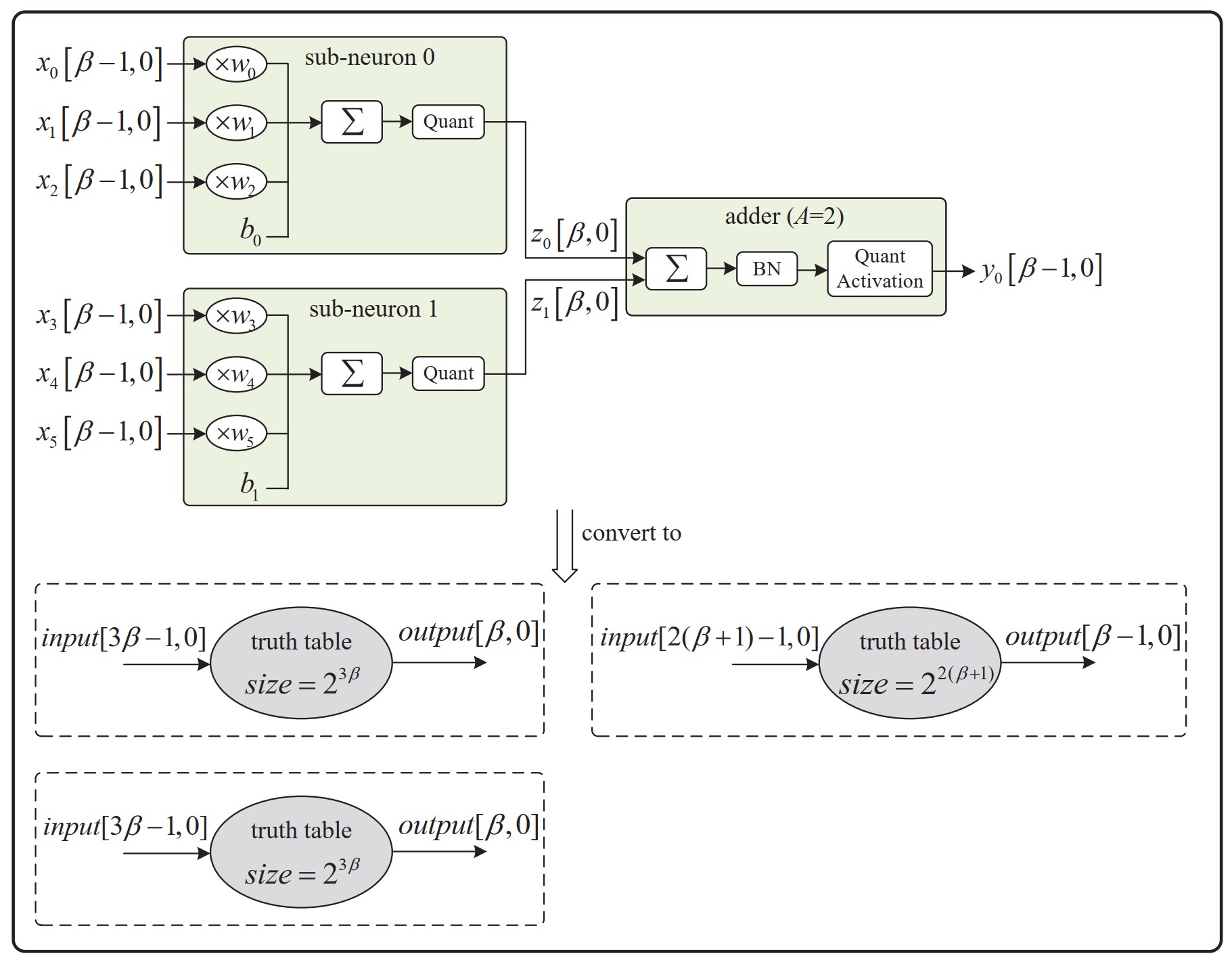}
        \caption{Architecture of a PolyLUT-Add neuron\cite{louPolyLUTAddFPGAbasedLUT}.}
        \label{polylut-add}
        \end{figure}

To mitigate the lookup table resource consumption of PolyLUT, PolyLUT-Add\cite{louPolyLUTAddFPGAbasedLUT} optimizes FPGA resource utilization by summing multiple PolyLUT sub-neurons, further reducing resource consumption and latency. As illustrated in Figure \ref{polylut-add}, the PolyLUT-Add neuron architecture employs an \(A\)-input adder to combine the outputs of \(A\) PolyLUT neurons (e.g., \(A = 2\) in the figure). For simplicity, assume the polynomial degree \(D = 1\). The sparse pruning method from PolyLUT is retained for inter-layer connections. The neuron output \(y_{0}\) is given by:

\begin{equation}
\label{polyadd}
\begin{aligned}
        y_{0} &= \sigma \left( \sum_{i = 0}^{AF-1} w_{i}x_{i} + b \right) \\
                &= \sigma \left( \sum_{a = 0}^{A-1} \left[ \sum_{i = 0}^{F-1} w_{(aF+i)}x_{(aF+i)} + b_{a} \right]  \right)
\end{aligned}
\end{equation}
where \(\sigma\) represents the activation function, \(w\) is the weight vector, and \(b\) is the bias parameter.

The LUT resource consumption between PolyLUT and PolyLUT-Add is compared through an example. Assuming a neuron input of \(F = 6\) and input bit width \(\beta = 2\), PolyLUT requires a lookup table size of \(2^{6 \beta} = 2^{12}\). In contrast, PolyLUT-Add, by combining \(A = 2\) PolyLUT sub-neurons, requires only \(2^{3 \beta} + 2^{3 \beta} + 2^{2 (\beta+1)} = 3 \times 2^{6}\), achieving a 95\% reduction in LUT resource consumption.

Although PolyLUT-Add reduces the lookup table size from \(\mathcal{O} (2^{\beta F A})\) to \(\mathcal{O} (A \times 2^{\beta F} + 2^{A(\beta+1)})\), it still experiences an exponential growth in lookup table size with increasing neuron inputs, leaving room for further improvement.

\subsection{NeuraLUT}

Compared to the multivariate polynomial function in PolyLUT, multilayer perceptrons (MLPs) are considered universal function approximators capable of capturing more complex features. Moreover, MLPs can be seamlessly embedded within the neurons of DNNs and trained through backpropagation. NeuraLUT \cite{andronicNeuraLUTHidingNeural2024} leverages this property by mapping entire sub-networks to a single L-LUT, rather than merely mapping individual neurons to L-LUTs as in previous approaches. This strategy enhances the expressive capacity of NeuraLUT, thereby reducing the number of circuit-level model layers, leading to lower resource consumption and latency.

\begin{figure}[t]
        \centering
        \includegraphics[width=2in]{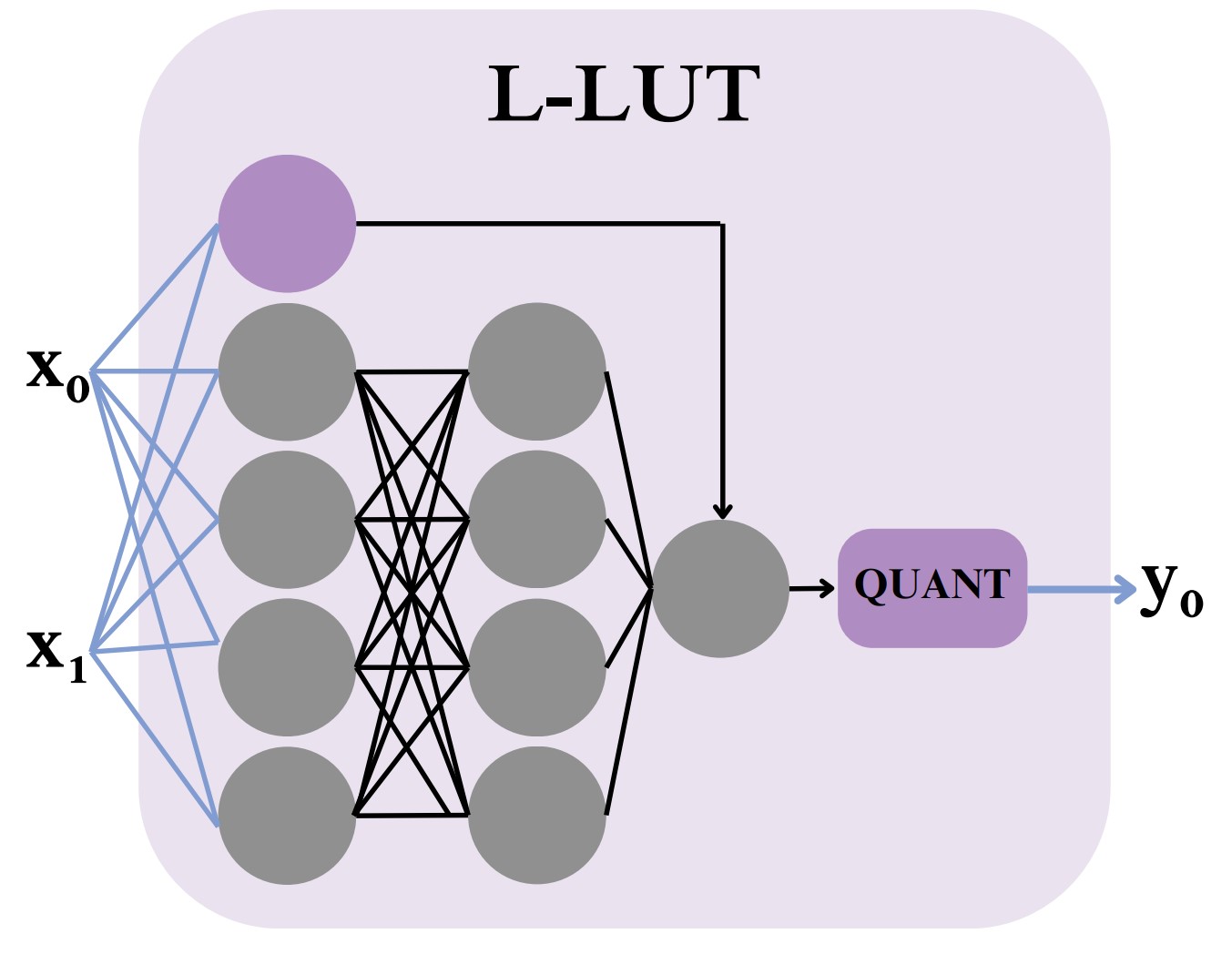}
        \caption{Architecture of a NeuraLUT neuron\cite{andronicNeuraLUTHidingNeural2024}.}
        \label{neura-lut}
        \end{figure}

Figure \ref{neura-lut} illustrates the neuron architecture of NeuraLUT, where MLPs are implemented within the L-LUT. The black lines indicate that MLP computations are performed in floating-point precision, meaning that internal MLP computations are not constrained by quantization or FPGA hardware limitations. The blue lines represent the quantized inputs and outputs of the L-LUT. Since the MLPs within the L-LUT may be relatively deep, training may suffer from the vanishing gradient problem. To enhance training stability, NeuraLUT employs skip connections, which allow gradients to bypass certain layers, thereby mitigating the vanishing gradient issue.

The detailed working principle of NeuraLUT is as follows. The sub-networks inside a single NeuraLUT neuron are denoted as $\mathcal{N}$, consisting of $L$ layers, where $S$ represents the number of layers skipped by the skip-connections. The overall function of the neuron, denoted as $f_{\mathcal{N}}$, consists of affine transformations $F_i$ and activation functions $\phi$. The symbol $\circ$ represents the composition of a layer's output serving as the input to the next layer. Thus, $f_{\mathcal{N}}$ can be expressed as:

\begin{equation}
\label{neura-func1}
f_{\mathcal{N}} = F_{\frac{L}{S}} \circ \phi \circ F_{\frac{L}{S} - 1} \circ \cdots \circ F_2 \circ \phi \circ F_1,
\end{equation}

The function of each layer $F_i(\mathbf{x})$ in the sub-network can be decomposed into the sum of $\hat{F}_i(\mathbf{x})$ and the skip-connection component $R_i(\mathbf{x})$:

\begin{equation}
\label{neura-func2}
F_i(\mathbf{x}) = \hat{F}_i(\mathbf{x}) + R_i(\mathbf{x}),
\end{equation}

where $\hat{F}_i(\mathbf{x})$ represents the function of an MLP, realized through a series of matrix multiplications $A_j$ and activation functions $\phi$. The transformation from layer $j = S(i-1)+1$ to layer $Si$ is given by:

\begin{equation}
\label{neura-func3}
\hat{F}_i(\mathbf{x}) = A_{Si} \circ \phi \circ A_{Si-1} \circ \cdots \circ \phi \circ A_{S(i-1)+1},
\end{equation}
where $A_j(\mathbf{x}) = \mathbf{W}_j \mathbf{x} + \mathbf{b}_j$, with $\mathbf{W}_j$ and $\mathbf{b}_j$ representing the weight and bias matrices, respectively. Compared to PolyLUT, the advantage of NeuraLUT lies in its ability to encapsulate more complex functions  within L-LUTs, thereby enhancing the model's expressive power. This allows NeuraLUT to achieve the same inference accuracy with fewer network layers or neurons, ultimately leading to lower resource consumption and reduced latency.

Although NeuraLUT employs fewer network layers, its training complexity increases with the depth of the sub-networks, and the size of the LUT still grows exponentially with the number of neuron inputs.

\subsection{AmigoLUT}

To address the issue of exponential LUT growth in LUT-based DNN models as neuron inputs increase, AmigoLUT \cite{wengGreaterSumIts2025} does not focus on optimizing the performance of a single model. Instead, it integrates multiple small LUT-based DNNs (such as LogicNets, PolyLUT, and NeuraLUT), ensuring that LUT resource consumption scales linearly with the number of models.

The main idea of AmigoLUT is that smaller and individually weak models, when combined, can outperform a single large model, much like how ensemble methods in machine learning often improve the overall performance by combining multiple weaker learners\cite{dongSurveyEnsembleLearning2020}. To determine the best method for combining multiple LUT-based models, the authors evaluate three well-known ensemble techniques: averaging, bagging, and AdaBoost. Averaging, which involves taking the mean of the outputs from all ensemble members, outperforms the other methods in their experiments. The key to averaging's success is that it promotes diversity among the ensemble members by allowing them to make slightly different predictions, thus compensating for the individual weaknesses of the models. 

\begin{figure}[t]
        \centering
        \includegraphics[width=3in]{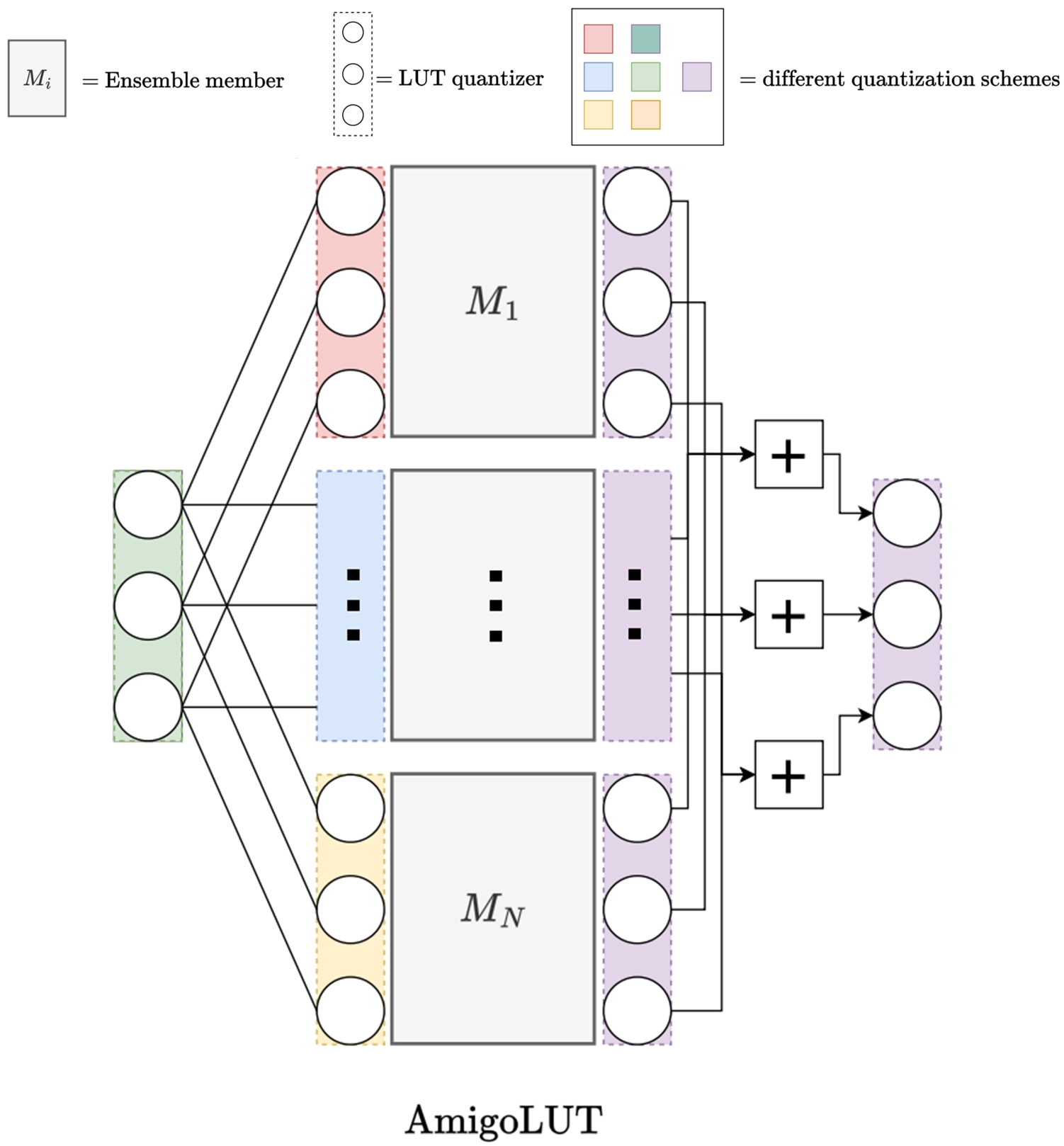}
        \caption{Architecture of AmigoLUT model\cite{wengGreaterSumIts2025}.}
        \label{amigolut}
\end{figure}

The key innovation of AmigoLUT is in how it handles quantization optimization, which is a major challenge for ensemble LUT-based networks because quantization introduces a challenge when ensemble members have different quantization schemes for their inputs and outputs. AmigoLUT solves this problem by using a hybrid quantization approach, as shown in Figure \ref{amigolut}. It first quantizes the input data to a low precision (e.g., 6 bits) using a shared quantizer across all ensemble members. This significantly reduces the need for multiple quantizers for each individual model, which would otherwise be computationally expensive. After this shared quantization step, the data is sent to each ensemble member, where each model can apply its own quantization to match the precision it was trained on. At the output stage, AmigoLUT applies an additional layer that quantizes the outputs of all ensemble members to the same precision before combining them. This ensures that the final results can be summed together efficiently while keeping the hardware cost manageable. This quantization scheme allows AmigoLUT to reduce FPGA resource usage while maintaining ensemble performance, offering a good balance between accuracy and resource efficiency.

Despite its success, AmigoLUT has a few limitations. First, the performance improvement is not unlimited. As the size of the ensemble increases, the accuracy gains start to diminish, particularly when larger base models are used. Additionally, the overall performance is still dependent on the base model architecture. A larger or more complex base model might offer better results initially, but ensembling can only provide limited benefits beyond a certain point. Moreover, while the quantization optimization significantly reduces resource overhead, it still requires careful tuning of the ensemble size and the quantization parameters.

\subsection{CompressedLUT}

\begin{figure}[b]
        \centering
        \includegraphics[width=3.5in]{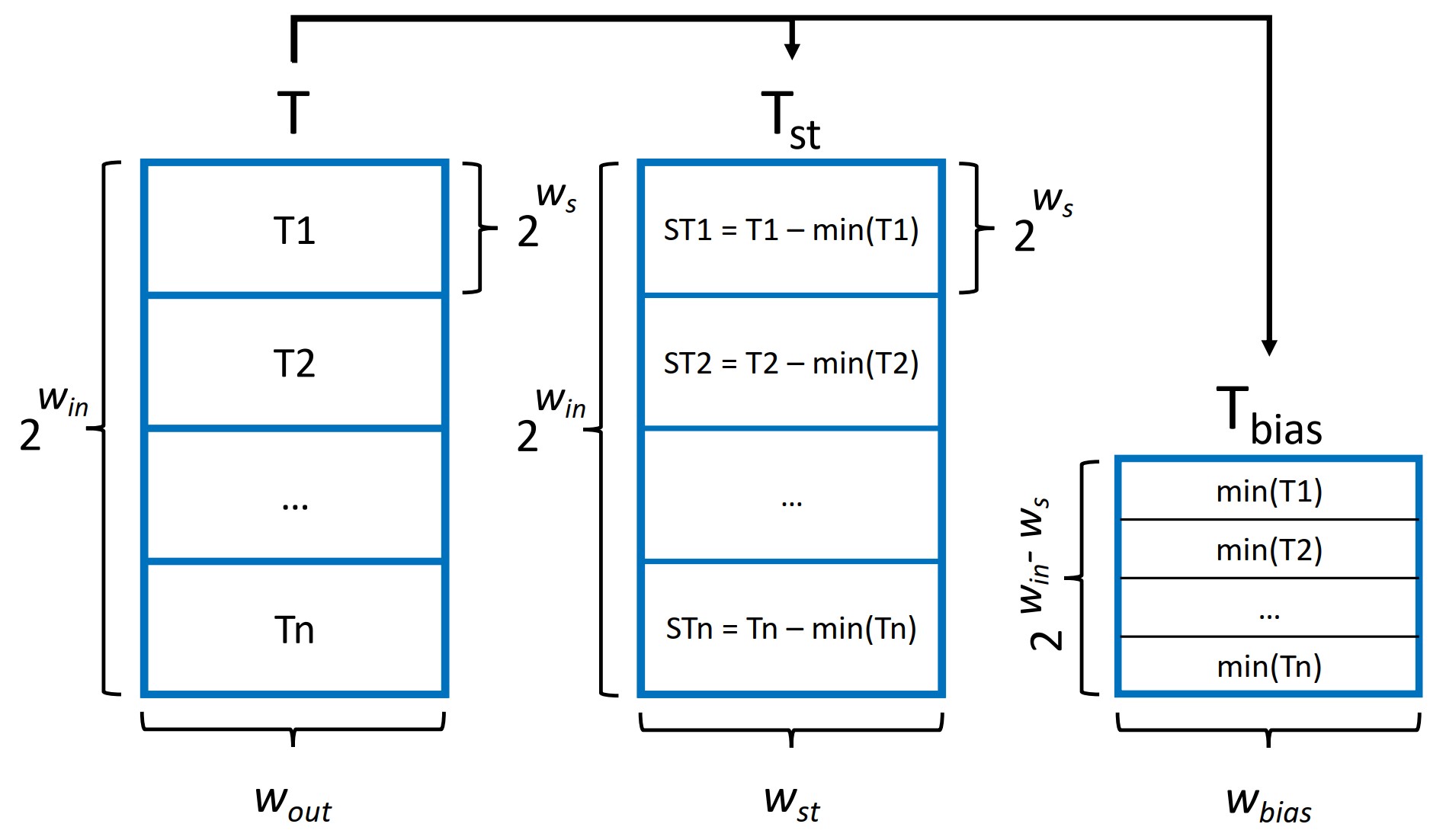}
        \caption{Decomposition of \(T\) into \( T_{\text{bias}} \) and \( T_{st} \)\cite{khataeiCompressedLUTOpenSource2024}.}
        \label{compressed-2}
        \end{figure}

To further reduce the significant LUT resource consumption of LUT-based DNN models, CompressedLUT \cite{khataeiCompressedLUTOpenSource2024} improves LUT utilization efficiency through bit-level compression techniques. The core objective of CompressedLUT is to minimize redundancy within LUTs, reducing storage requirements while maintaining inference accuracy. This compression method enables LUT-based models to scale effectively within FPGA resource constraints, thereby improving hardware efficiency.

CompressedLUT employs bit-level compression techniques consisting of three main components: lookup table decomposition, self-similarity exploitation, and higher-bit compression. The method first decomposes the lookup table $T$ into a bias table $T_{\text{bias}}$ and sub-tables $T_{st}$. Figure \ref{compressed-2} illustrates this decomposition process, where $T_{\text{bias}}$ stores the minimum values within the sub-tables, while $T_{st}$ stores the residual values after subtracting the minimum. To further compress $T_{st}$, CompressedLUT identifies self-similarities within the lookup table. By constructing a similarity matrix $SM$, it detects sub-tables that are identical or derivable through arithmetic right shifts. If two sub-tables $ST_i$ and $ST_j$ satisfy:

\begin{equation}
\label{compress-1}
SM_{i,j} = 1 \iff \exists t : \forall k, \ ST_i[k] \gg t = ST_j[k],
\end{equation}
then $ST_j$ can be obtained by right-shifting $ST_i$. 

\begin{figure}[t]
        \centering
        \includegraphics[width=3in]{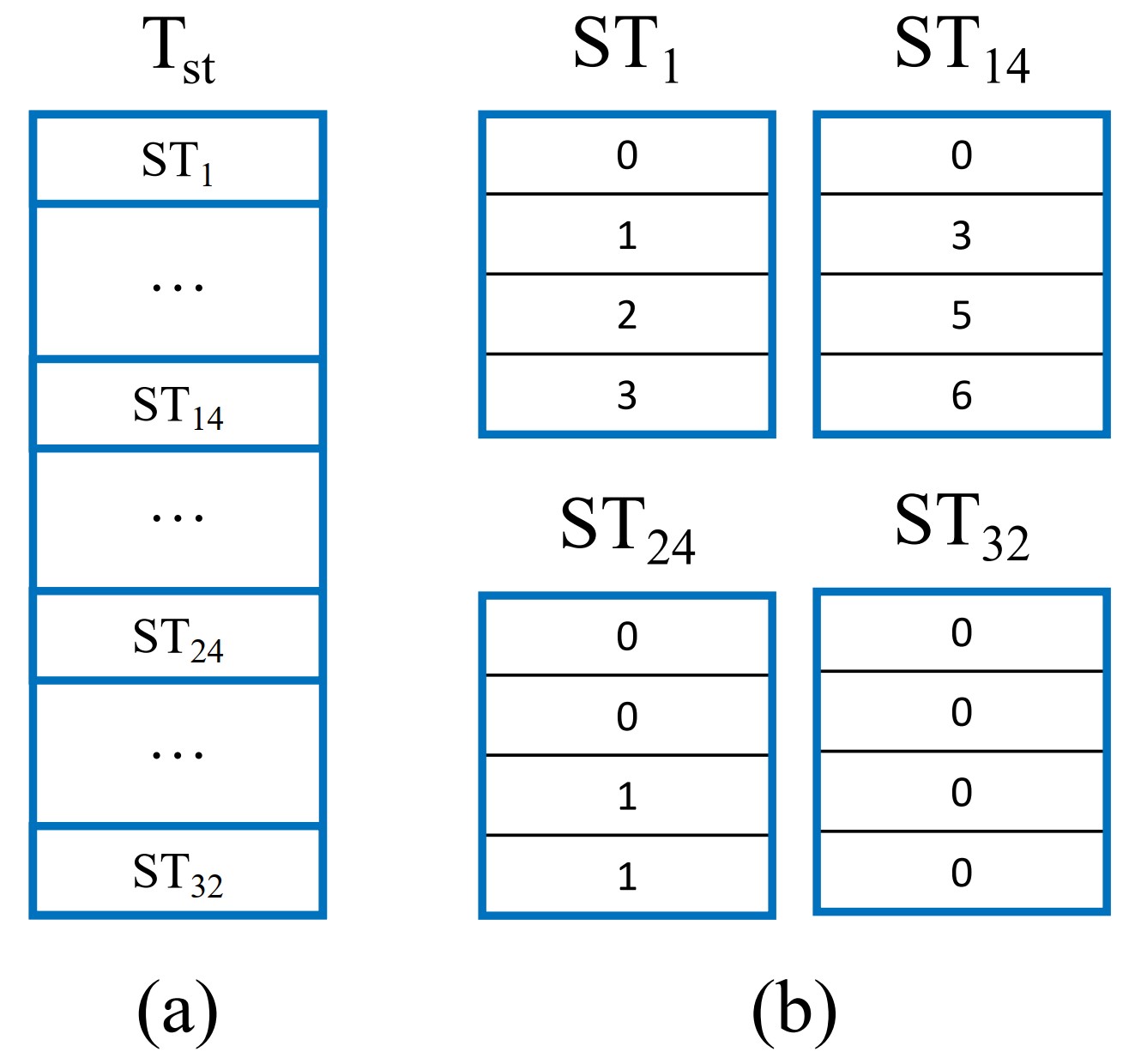}
        \caption{Examples of self-similarities in sub-tables\cite{khataeiCompressedLUTOpenSource2024}.}
        \label{compressed-3}
        \end{figure}

Figure \ref{compressed-3} provides an example of LUT self-similarities, where $T_{st}$ consists of 32 sub-tables, each containing four elements. As shown in Figure \ref{compressed-2}(b), sub-table $ST_{14}$ can be right-shifted by one bit to obtain $ST_1$, by two bits to obtain $ST_{24}$, and by three bits to obtain $ST_{32}$. In this case, $ST_{14}$ is designated as a unique sub-table, while the others can be derived through right shifts. The goal of CompressedLUT is to find the minimal set of unique sub-tables $T_{ust}$ and use an index table $T_{idx}$ and a right shift information table $T_{rsh}$ to reconstruct all other sub-tables, thereby maximizing LUT reuse and minimizing resource consumption.

\begin{figure}[b]
        \centering
        \includegraphics[width=3.5in]{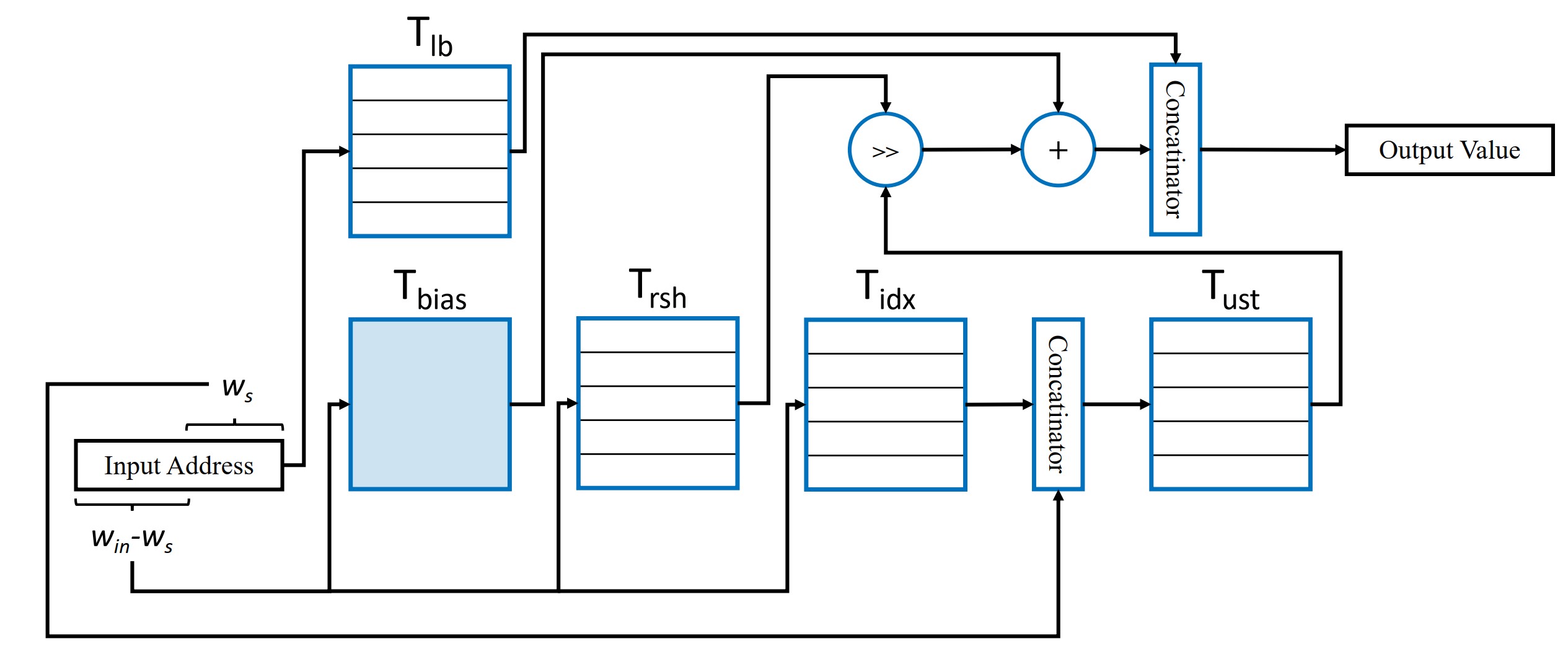}
        \caption{Overall architecture of CompressedLUT method\cite{khataeiCompressedLUTOpenSource2024}.}
        \label{compressed}
        \end{figure}

For LUTs with a large dynamic range and strong local variations, CompressedLUT applies higher-bit compression techniques. The LUT $T$ is decomposed into a high-bit table $T_{hb}$ and a low-bit table $T_{lb}$, where $T_{lb}$ stores raw values without compression, while $T_{hb}$ undergoes decomposition and self-similarity detection for compression. The overall architecture of CompressedLUT method is illustrated in Figure \ref{compressed}, where the same architecture is embedded in $T_{\text{bias}}$. The final LUT output is reconstructed by concatenating $T_{hb}$ and $T_{lb}$, formulated as:

\begin{equation}
\label{compress-0}
T[x] = (T_{ust}[{T_{idx}[x_{hb}],x_{lb}}] \gg T_{rsh}[x_{hb}]) + T_{bias}[x_{hb}].
\end{equation}

Despite its effectiveness in LUT resource optimization, CompressedLUT still faces several challenges. As DNN scales increase, certain computational behaviors may exhibit randomness or lack discernible patterns, making LUT internal functions highly complex. This reduces self-similarity, thereby limiting the efficiency of CompressedLUT’s pattern recognition methodology.

\subsection{ReducedLUT}

To address the inefficiency of CompressedLUT when scaling up DNNs, ReducedLUT\cite{cassidyReducedLUTTableDecomposition2024} further optimizes the LUT compression method based on CompressedLUT. It introduces a technique specifically designed to reduce LUT resource consumption in LUT-based DNNs. Unlike CompressedLUT, which primarily focuses on reducing bitwidth, ReducedLUT leverages "don't care" values, which are input values that do not affect the output. By utilizing these don't care values, ReducedLUT aims to map more input values to existing sub-tables, thereby reducing the number of unique sub-tables $T_{ust}$ and further minimizing the LUT size.

\begin{figure}[b]
\centering
\includegraphics[width=2.5in]{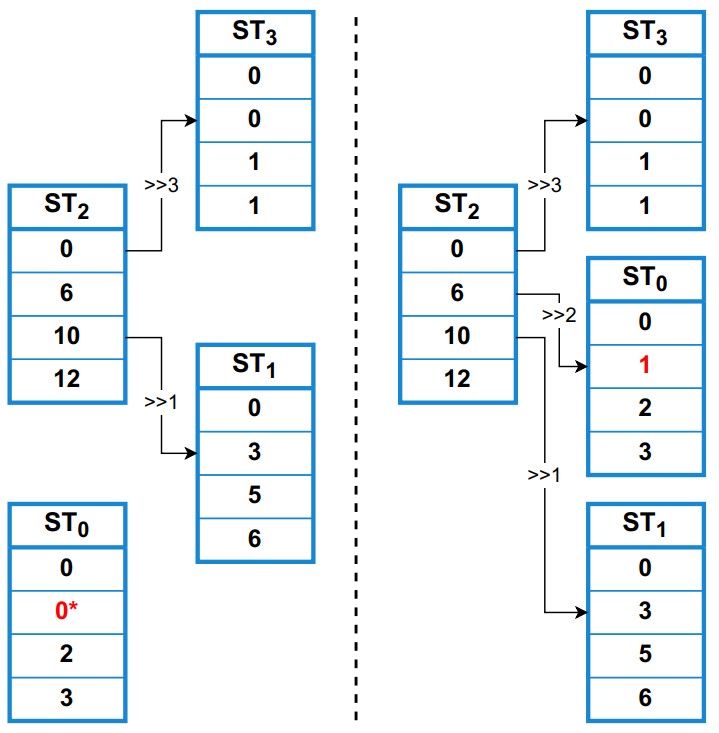}
\caption{Don't care usage example\cite{cassidyReducedLUTTableDecomposition2024}.}
\label{reducedlut}
\end{figure}

For instance, as shown in Figure \ref{reducedlut}, the initial table $T$ can be decomposed into four sub-tables: $ST_0$, $ST_1$, $ST_2$, and $ST_3$. Without incorporating don't care values, only \( ST_1 \) and \( ST_3 \) can be reconstructed through \( ST_2 \). However, by introducing a don't care value in sub-table \( ST_0 \), additional compression can be achieved—simply setting its second element to 1 allows \( ST_0 \) to shift right by two positions and be reconstructed through \( ST_2 \).

ReducedLUT infers from the training data and marks input combinations that do not appear in the training set as don't care values. Since these conditions are absent from the training data, their corresponding output values can be freely adjusted without affecting training accuracy. During the LUT compression process, ReducedLUT follows the compression workflow of CompressedLUT: first, decomposing the LUT into multiple sub-tables; then, computing a similarity matrix \(SM\) for all sub-tables; next, summing the columns of the SM to obtain a similarity vector \(SV\); and finally, ranking sub-tables based on the similarity vector, prioritizing optimization for the least dependent sub-tables. The key difference is that ReducedLUT can modify the values of don't care conditions, enabling more sub-tables to be reconstructed using existing ones. Additionally, it employs a Boolean mask to protect already optimized sub-tables from redundant modifications that might affect the final structure. ReducedLUT adopts an iterative optimization strategy, repeating the optimization process under various sub-table sizes and higher-bit compression configurations to identify the optimal compression scheme. This approach significantly reduces LUT resource consumption while maintaining DNN inference accuracy.

Compared to CompressedLUT, ReducedLUT optimizes the LUT decomposition process by identifying unobserved patterns in DNN training data, which are often ignored by CompressedLUT. Fundamentally, ReducedLUT leverages the domain characteristics of DNN training data to enhance pattern recognition, thereby improving LUT self-similarity and making LUT decomposition more efficient.

\subsection{DiffLogicNet}
\begin{figure*}[t]
        \centering
        \includegraphics[width=6in]{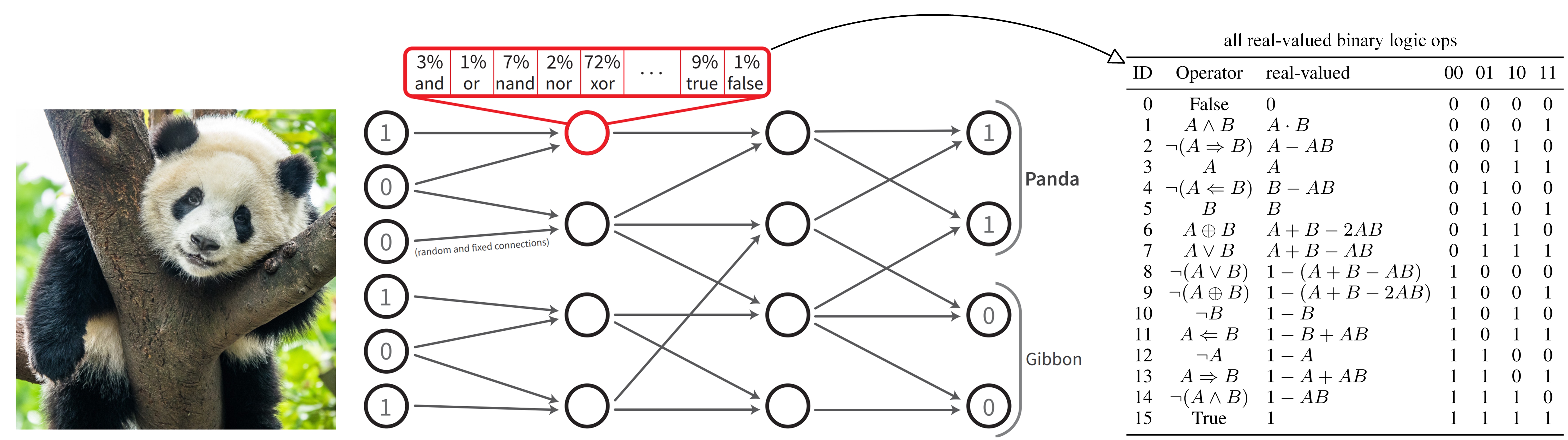}
        \caption{Overview of the DiffLogicNet \& list of all real-valued binary logic ops\cite{petersenDeepDifferentiableLogic2022}.}
        \label{difflogicnets}
        \end{figure*}

The \ac{VC}-dimension is a mathematical concept used to measure the capacity of a hypothesis class. Specifically, given a hypothesis class \( \mathcal{H} \), its VC-dimension is defined as the size of the largest set of points that can be "shattered" by \( \mathcal{H} \). In other words, the VC-dimension of \( \mathcal{H} \) is the largest integer \( d \) for which there exists a set of \( d \) points that can realize all possible binary classification label combinations under \( \mathcal{H} \). For an \( n \)-input LUT, its VC-dimension is \( 2^n \)\cite{carneiroExactVCDimension2019}, meaning it can represent all possible \( n \)-input binary functions. In contrast, an \( n \)-input DNN neuron can only implement \( n+1 \) classification label combinations, resulting in a VC-dimension of \( n+1 \).

While previous LUT-based DNN architectures primarily focused on converting pre-trained models into LUTs for efficient inference, recent advancements have introduced a paradigm shift: directly training LUT-based \ac{WNNs} models using differentiable methods. WNNs avoid traditional weighted connections in favor of a multiplication-free approach, using binary-valued LUTs, or "RAM nodes", to dictate neuronal activity\cite{bacellarDifferentiableWeightlessNeural2024}. This innovative approach eliminates the need for post-training conversion and allows the LUTs themselves to be updated dynamically via gradient-based optimization, leveraging the computational capabilities of LUTs more. The success of these methods lies in their ability to directly train $n$-input LUTs, which have a VC-dimension of $2^{n}$, enabling the construction of smaller and more efficient models. In contrast, previous approaches trained $n$-input DNN neurons with a VC-dimension of $n+1$ and then converted them into LUTs merely as a speedup mechanism. This transformation led to larger and less efficient models, as it did not fully exploit the representational capacity of LUTs.


DiffLogicNet\cite{petersenDeepDifferentiableLogic2022} proposed an approach to learning multi-layer networks exclusively composed of binary logic which are exactly the same as a WNN with two-input LUTs (LUT-2s), as illustrated in Figure \ref{difflogicnets}. In this model, an input binary vector is processed through multiple layers of binary logic nodes. These nodes are randomly connected and remain static, ultimately leading to a final summation determining the output class score. For training these networks via gradient descent, DiffLogicNet proposes a method where binary values are relaxed into probabilities. This is achieved by considering all possible binary logic functions which is listed in Figure \ref{difflogicnets} and then applying a softmax function to create a probability distribution over these logic functions. 

DiffLogicNet's methodology revolves around defining each neuron as a probabilistic mixture of logic gates. Given binary inputs \( a_1, a_2 \), the output is computed as:

\begin{equation}
\label{DWN-1}
a' = \sum_{i=0}^{15} p_i f_i(a_1, a_2) = \sum_{i=0}^{15} \frac{e^{w_i}}{\sum_j e^{w_j}} f_i(a_1, a_2),
\end{equation}
where \( f_i \) denotes one of the 16 logic operations (e.g., AND, OR, XOR), and \( p_i \) represents a probability distribution learned via softmax.

This relaxation enables differentiability, allowing standard backpropagation to optimize the network structure. However, the cost of this method scales double-exponentially ($\mathcal{O}$($2^{2^n}$)) with the number of inputs to each LUT (e.g., 16 different functions for 2 variables), requiring an astonishing 18.4 quintillion parameters to represent a single LUT-6s. Additionally, it relies on pseudorandom connections, leading to suboptimal network structures, and uses popcounts for activation, which significantly increases hardware overhead. Finally, its binary LUT-based architecture prevents the use of traditional DNN regularization techniques, making it prone to overfitting.

\subsection{DWN}

To overcome the limitations of DiffLogicNet, DWN\cite{bacellarDifferentiableWeightlessNeural2024} introduces a series of enhancements that systematically address its drawbacks. DWN employs the Extended Finite Difference (EFD) method to efficiently compute the gradients of LUTs. This method approximates the LUT gradient using the mathematical formulation in Equation \ref{DWN-2}, eliminating the need to store all possible logic mappings. For a function \( A: \mathbb{R}^{2^n} \times \{0,1\}^n \to \mathbb{R} \) that maps binary input \( a \) to a LUT output, the EFD gradient approximation is given by:

\begin{equation}
        \label{DWN-2}
        \frac{\partial A}{\partial a_j}(U, a) = \sum_{k\in\{0,1\}^n} (-1)^{(1-k_j)} A(U, k) \cdot H(k, a, j) + 1,
        \end{equation}
where \( H(k, a, j) \) computes the Hamming distance between LUT addresses. This method leverages stored LUT values to compute gradients without explicitly storing all possible logic mappings, reducing the parameter storage requirement from \( \mathcal{O}(2^{2^n}) \) in DiffLogicNet to \( \mathcal{O}(2^n) \), thereby making large-scale LUTs feasible.

\begin{figure}[b]
\centering
\includegraphics[width=3in]{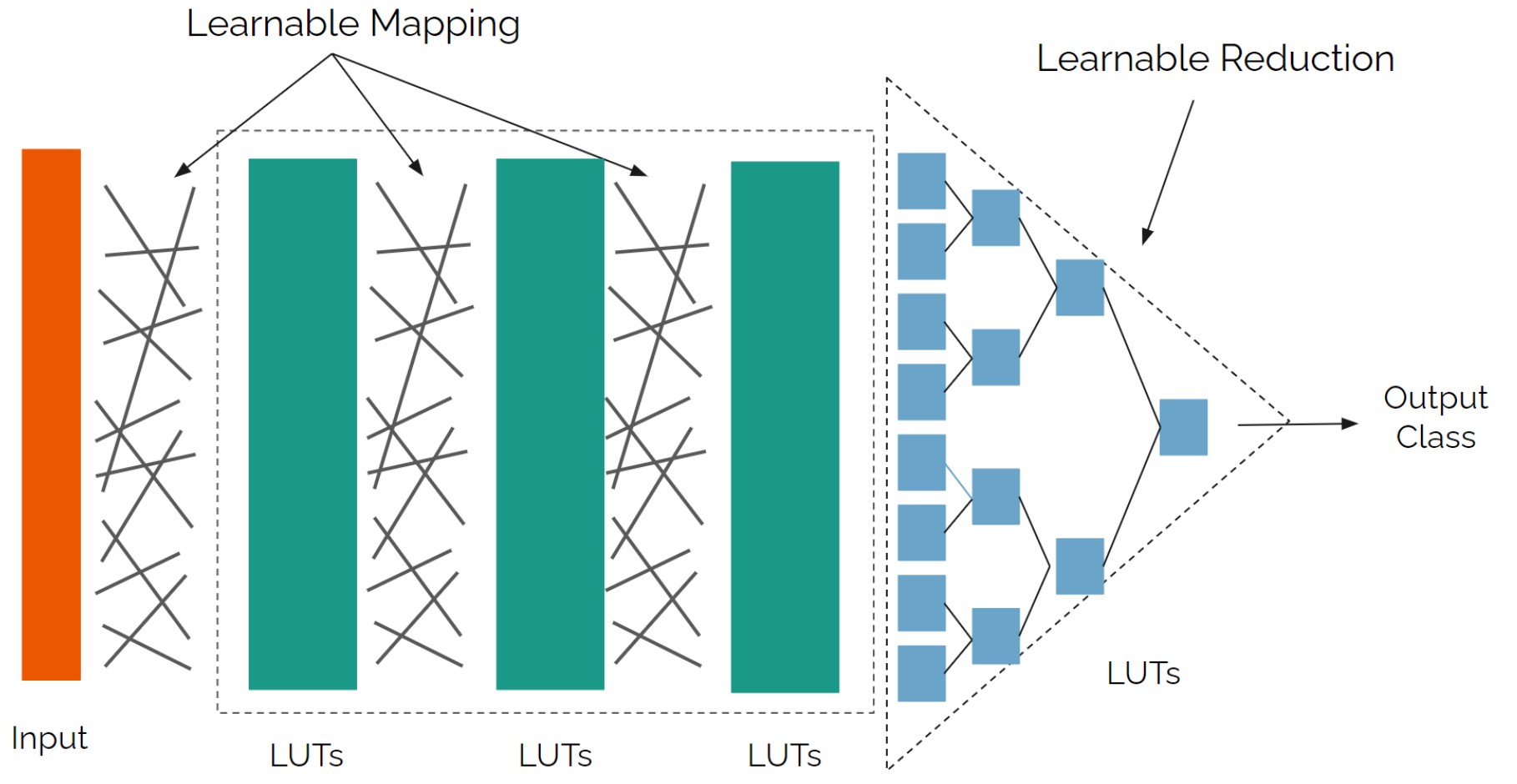}
\caption{Learnable Mapping \& Learnable Reduction in DWNs\cite{bacellarDifferentiableWeightlessNeural2024}.}
\label{DWN}
\end{figure}

Unlike DiffLogicNet, which relies on static pseudo-random mappings, DWN employs learnable mapping, allowing the network to dynamically optimize LUT connections, thereby improving adaptability and efficiency. To mitigate the area overhead introduced by the popcount computations in DiffLogicNet, DWN incorporates learnable reduction, utilizing pyramidal LUT layers to optimize activation computations. Figure \ref{DWN} illustrates DWN’s learnable mapping and learnable reduction techniques.

To address the overfitting issues in DiffLogicNet, DWN employs spectral regularization by constraining the spectral norm of LUT functions to enhance generalization. For an input-output mapping \( f: \{0,1\}^n \to \mathbb{R} \), its spectral norm is defined as:

\begin{equation}
        \label{DWN-3}
        \text{specnorm}(L) = \| L C \|_2, \quad C_{ij} = \frac{1}{2^n} \prod_{a \in \{b \mid i_b=1\}} (2j_a - 1),
        \end{equation}
where \( L \) represents the data matrix storing LUT values, containing either binary or computed numerical values. \( C_{ij} \) is a precomputed coefficient matrix that constructs the spectral norm of LUT computations. The product term \( \prod_{a \in \{b \mid i_b=1\}} (2j_a - 1) \) reflects the relationship between LUT input binary patterns and stored values, while the normalization factor \( \frac{1}{2^n} \) ensures numerical stability. By constraining the \( L_2 \) norm of \( L \) after transformation by \( C \), DWN prevents LUTs from excessively relying on specific input patterns, thereby improving model generalization.

Although DWN effectively leverages the computational power of LUTs, its overhead still grows exponentially with LUT size. Furthermore, while EFD enables gradient-based training, it remains an approximation method, which may lead to suboptimal learning dynamics.

\subsection{TreeLUT}

TreeLUT\cite{khataeiTreeLUTEfficientAlternative2025} presents a significant advancement over previous LUT-based DNN architectures by leveraging Gradient Boosted Decision Trees (GBDTs) as an alternative to DNNs for inference acceleration on FPGAs. Unlike previous LUT-based DNNs, which relies on LUT-based approximations of DNNs neurons and requires extensive quantization that can degrade accuracy, TreeLUT inherently avoids the computational complexity of DNNs by using decision trees, which are natively interpretable and efficiently mapped onto LUTs. Furthermore, TreeLUT employs a novel quantization strategy that optimizes decision thresholds and leaf values before training, eliminating the need for post-training transformations.

The methodology of TreeLUT is centered around the efficient quantization and hardware mapping of GBDTs. Given an input feature vector \( X \), a trained GBDT model consists of \( M \) decision trees, each performing a series of comparisons at internal nodes and outputting continuous leaf values. The final prediction is computed as the sum of all tree outputs:  

\begin{equation}
\label{tree-1}
F(X) = f_0 + \sum_{m=1}^{M} f_m(X),
\end{equation}
where \( f_0 \) is an initial prediction score, and \( f_m(X) \) is the output of the \( m \)-th decision tree. In binary classification, the probability is obtained using a sigmoid function, while in multiclass classification, a one-vs-all strategy followed by a softmax function is applied. To make this structure hardware-friendly, TreeLUT introduces pre-training quantization of input features and post-training quantization of leaf values, ensuring that all computations remain integer-based. The feature quantization step maps continuous input values into discrete levels:  

\begin{equation}
\label{tree-2}
X_{\text{quantized}} = \text{round} \left( X_{\text{normalized}} \times (2^{w_{\text{feature}}} - 1) \right),
\end{equation}
where \( w_{\text{feature}} \) is a quantization hyperparameter. Similarly, leaf values are adjusted to fit within a reduced bit-width: 

\begin{equation}
\label{tree-3}
F'(X) = b + \sum_{m=1}^{M} f'_m(X),
\end{equation}
where \( f'_m(X) = f_m(X) - \min(f_m) \) ensures non-negative values, and \( b \) is a bias term. 

\begin{figure}[t]
        \centering
        \includegraphics[width=3in]{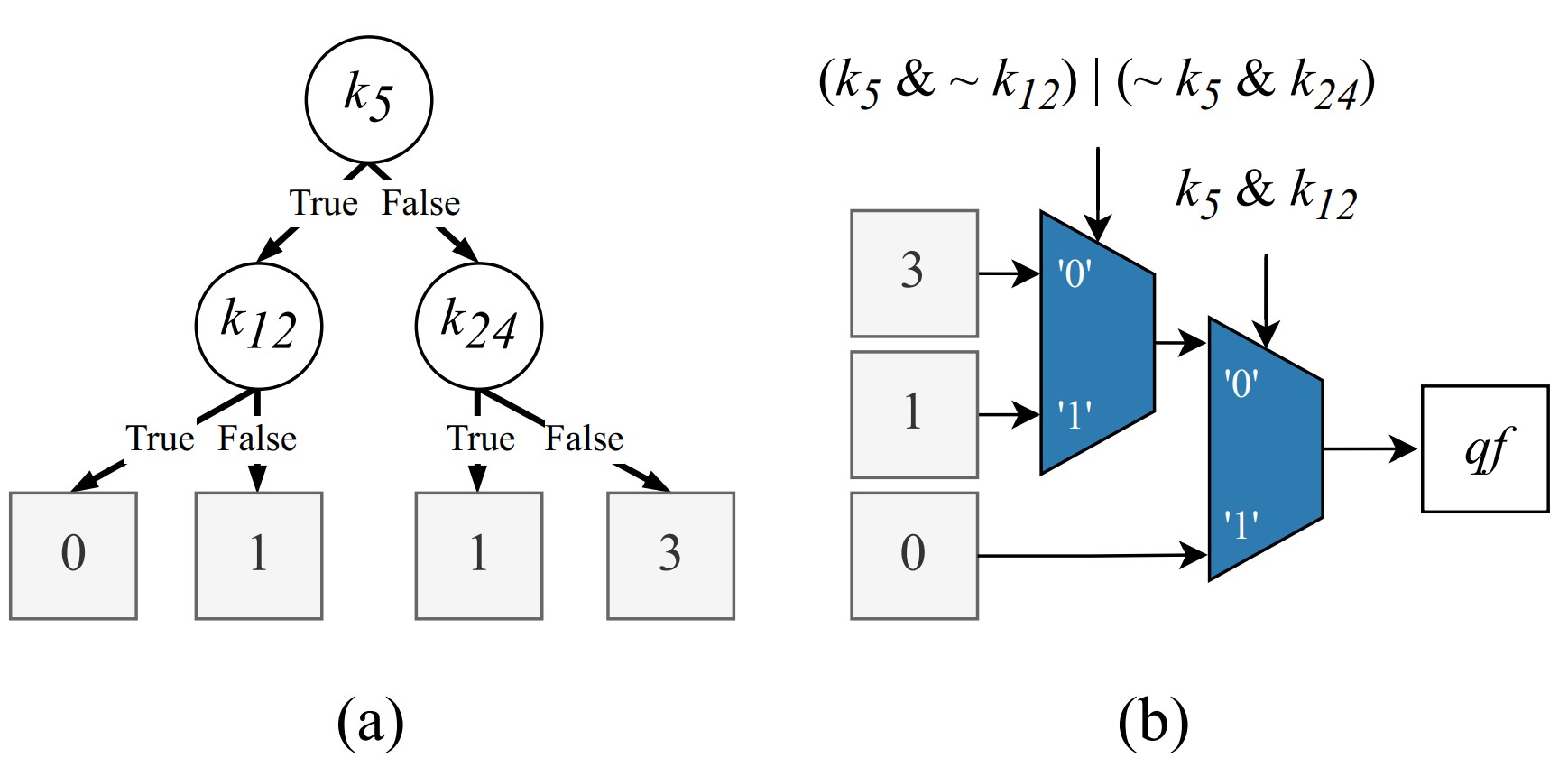}
        \caption{The architecture of a decision tree\cite{khataeiTreeLUTEfficientAlternative2025}.}
        \label{treefig-1}
\end{figure}

\begin{figure}[t]
\centering
\includegraphics[width=3in]{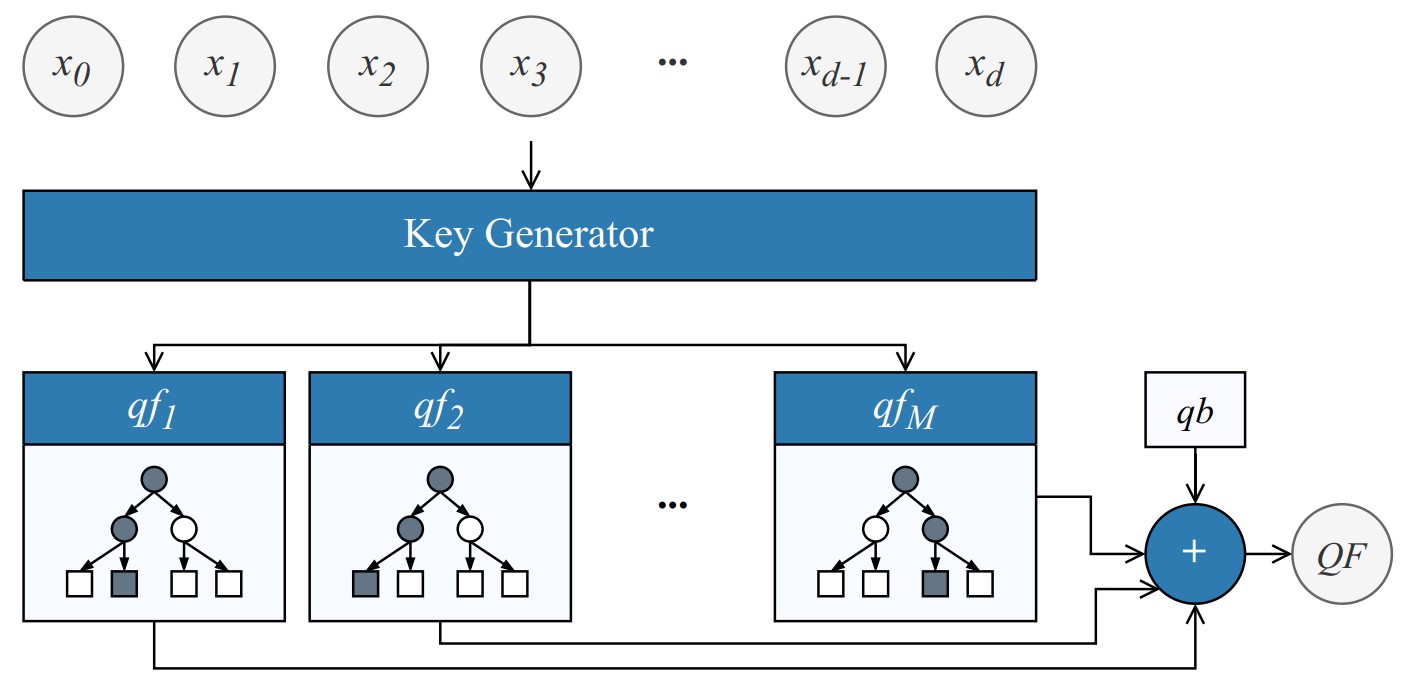}
\caption{The overall TreeLUT architecture for binary classification tasks\cite{khataeiTreeLUTEfficientAlternative2025}.}
\label{treefig-2}
\end{figure}

To further optimize hardware implementation, TreeLUT represents decision trees using multiplexer-based logic, where each path from root to leaf is encoded as a Boolean function. As illustrated in Figure \ref{treefig-1}, an adder tree structure efficiently aggregates the tree outputs, minimizing critical path delay. Figure \ref{treefig-2} demonstrates the overall TreeLUT architecture for binary classification tasks. The TreeLUT hardware architecture consists of three key layers: a key generator, which performs parallel comparisons between input features and decision thresholds; a decision tree layer, which represents trees as Boolean expressions implemented using LUTs; and an adder tree, which accumulates outputs efficiently.

Despite its efficiency, TreeLUT has several limitations that affect its scalability and applicability. While decision trees are inherently well-suited for tabular data, they may struggle with highly unstructured inputs such as images or sequential data, where DNNs excel due to their ability to learn hierarchical feature representations. Furthermore, TreeLUT's lack of adaptive learning mechanisms means that unlike DNNs, which can be fine-tuned via backpropagation, its models must be entirely retrained from scratch when updated.

\section{Experimental Results Comparison}

\begin{table*}
\begin{center}
\caption{Comparison of LUT-based DNNs and DTs in terms of accuracy and hardware costs. The results were quoted directly from their original papers.}
\label{tabular}

\resizebox{\textwidth}{!}{  

\begin{threeparttable}
\begin{tabular}{c|l|c|c|c|c|c|c|c|c|c}
\hline
\multirow{2}{*}{Dataset} & \multirow{2}{*}{\centering Method} & \multirow{2}{*}{Model} & \multirow{2}{*}{Accuracy} & \multicolumn{7}{c}{Hardware Costs} \\ 
\cline{5-11} 
& & & & LUT & FF & DSP & BRAM & Fmax (MHz) & Latency (ns) & Area $\times$ Delay \\ 
\hline
\multirow{10}{*}{MNIST} & TreeLUT (I)\cite{khataeiTreeLUTEfficientAlternative2025} & DT & 97\% & 4,478 & 597 & 0 & 0 & 791 & 2.5 & 1.12e+4  \\
        & TreeLUT (II)\cite{khataeiTreeLUTEfficientAlternative2025} & DT & 96\% & 3,499 & 759 & 0 & 0 & 874 & \textbf{2.3} & 8.05e+3  \\
        & AmigoLUT-LogicNet-XS (2 models)\cite{wengGreaterSumIts2025} & DNN & 94.7\% & 9,711 & 9,047 & 0 & 0 & 569 & 12.3 & 1.2e+4  \\
        & AmigoLUT-NeuraLUT (4 models)\cite{wengGreaterSumIts2025} & DNN & 95.5\% & 16,081 & 13,292 & 0 & 0 & \textbf{925} & 7.6 & 1.19e+4  \\
        & ReducedLUT (HDR-5L)\cite{cassidyReducedLUTTableDecomposition2024} & DNN & 95.7\% & 47,484 & — & 0 & 0 & 295 & 17 & 8.07e+5  \\
        & DWN (sm)\cite{bacellarDifferentiableWeightlessNeural2024} & DNN & 97\% & \textbf{692} & \textbf{422} & 0 & 0 & 827 & 2.4 & \textbf{1.66e+3}  \\
        & DWN (md)\cite{bacellarDifferentiableWeightlessNeural2024} & DNN & 97.9\% & 1,413 & 1,143 & 0 & 0 & 827 & 3.6 & 5.08e+3  \\
        & DWN (lg)\cite{bacellarDifferentiableWeightlessNeural2024} & DNN & \textbf{98.3\%} & 4,082 & 3,385 & 0 & 0 & 827 & 6.0 & 2.45e+4  \\
        & PolyLUT-Add (HDR-Add2)\cite{louPolyLUTAddFPGAbasedLUT} & DNN & 96\% & 14,810 & 2,609 & 0 & 0 & 625 & 10 & 1.48e+5  \\
        & NeuraLUT (HDR-5L)\cite{andronicNeuraLUTHidingNeural2024} & DNN & 96\% & 54,798 & 3,757 & 0 & 0 & 431 & 12 & 6.58e+5  \\
        & PolyLUT (HDR)\cite{andronicPolyLUTLearningPiecewise2023b} & DNN & 96\% & 70,673 & 4,681 & 0 & 0 & 378 & 16 & 1.13e+6  \\
        & PolyLUT (HDR)\tnote{1} & DNN & 96\% & 61,500 & 4,237 & 0 & 0 & 294 & 20 & 1.13e+6  \\
\hline
\multirow{9}{*}{JSC-H} & TreeLUT (I)\cite{khataeiTreeLUTEfficientAlternative2025} & DT & 76\% & 2,234 & 347 & 0 & 0 & 735 & 2.7 & 6.03e+3  \\
        & TreeLUT (II)\cite{khataeiTreeLUTEfficientAlternative2025} & DT & 75\% & 796 & \textbf{74}& 0 & 0 & \textbf{887} & \textbf{1.1} & \textbf{8.76e+2}  \\
        & AmigoLUT-NeuraLUT-S (32 models)\cite{wengGreaterSumIts2025} & DNN & 74.4\% & 42,742 & 4,717 & 0 & 0 & 520 & 9.6 & 4.1e+5  \\
        & ReducedLUT (JSC-5L)\cite{cassidyReducedLUTTableDecomposition2024} & DNN & 74.9\% & 58,409 & — & 0 & 0 & 302 & 17 & 9.92e+5  \\
        & DWN (md)\cite{bacellarDifferentiableWeightlessNeural2024} & DNN & 75.6\% & \textbf{720} & 457 & 0 & 0 & 827 & 3.6 & 2.59e+3  \\
        & DWN (lg)\cite{bacellarDifferentiableWeightlessNeural2024} & DNN & \textbf{76.3\%} & 4,972 & 3,305 & 0 & 0 & 827 & 7.3 & 3.63e+4  \\
        & PolyLUT-Add (JSC-XL-Add2)\cite{louPolyLUTAddFPGAbasedLUT} & DNN & 75\% & 36,484 & 1,209 & 0 & 0 & 315 & 16 & 5.84e+5  \\
        & NeuraLUT (JSC-5L)\cite{andronicNeuraLUTHidingNeural2024} & DNN & 75\% & 92,357 & 4,885 & 0 & 0 & 368 & 14 & 1.29e+6  \\
        & PolyLUT (JSC-XL)\cite{andronicPolyLUTLearningPiecewise2023b} & DNN & 75\% & 236,541 & 2,775 & 0 & 0 & 235 & 21 & 4.97e+6  \\
        & PolyLUT (JSC-XL)\tnote{1} & DNN & 75\% & 168,746 & 8,793 & 0 & 0 & 204 & 24 & 4.97e+6  \\
\hline
\multirow{9}{*}{JSC-L} & AmigoLUT-NeuraLUT-XS (4 models)\cite{wengGreaterSumIts2025} & DNN & 71.1\% & 320 & 482 & 0 & 0 & 1445 & 3.5 & 1.12e+3 \\
        & AmigoLUT-NeuraLUT-XS (16 models)\cite{wengGreaterSumIts2025} & DNN & 72.9\% & 1,243 & 1,240 & 0 & 0 & 1008 & 5.0 & 6.21e+3  \\
        & ReducedLUT (JSC-2L)\cite{cassidyReducedLUTTableDecomposition2024} & DNN & 72.5\% & 2,786 & — & 0 & 0 & 408 & 4.9 & 1.36e+4  \\
        & DWN (sm)\cite{bacellarDifferentiableWeightlessNeural2024} & DNN & 71.1\% & \textbf{20} & \textbf{22} & 0 & 0 & \textbf{3030} & \textbf{0.6} & \textbf{1.30e+1}  \\
        & DWN (sm)\cite{bacellarDifferentiableWeightlessNeural2024} & DNN & \textbf{74\%} & 110 & 72 & 0 & 0 & 1094 & 1.5 & 1.65e+2  \\
        & PolyLUT-Add (JSC-M Lite-Add2)\cite{louPolyLUTAddFPGAbasedLUT} & DNN & 72\% & 895 & 189 & 0 & 0 & 750 & 4 & 3.58e+3  \\
        & NeuraLUT (JSC-2L)\cite{andronicNeuraLUTHidingNeural2024} & DNN & 72\% & 4,684 & 341 & 0 & 0 & 727 & 3 & 1.40e+4  \\
        & PolyLUT (JSC-M Lite)\cite{andronicPolyLUTLearningPiecewise2023b} & DNN & 72\% & 12,436 & 773 & 0 & 0 & 646 & 5 & 6.22e+4  \\
        & PolyLUT (JSC-M Lite)\tnote{1} & DNN & 72\% & 10,482 & 629 & 0 & 0 & 465 & 7 & 7.34e+4  \\
        & LogicNets\cite{umurogluLogicNetsCoDesignedNeural2020} & DNN & 72\% & 37,900 & 810 & 0 & 0 & 384 & 13 & 4.93e+5  \\
\hline
\multirow{5}{*}{NID} & TreeLUT (I)\cite{khataeiTreeLUTEfficientAlternative2025} & DT & \textbf{93\%} & 345 & 33 & 0 & 0 & 681 & 1.5 & 5.18e+2  \\
        & TreeLUT (II)\cite{khataeiTreeLUTEfficientAlternative2025} & DT & 92\% & \textbf{89} & \textbf{19} & 0 & 0 & \textbf{1,047} & \textbf{1.0} & \textbf{8.90e+1} \\
        & PolyLUT-Add (NID-Add2)\cite{louPolyLUTAddFPGAbasedLUT} & DNN & 92\% & 1,649 & 830 & 0 & 0 & 620 & 8 & 1.32e+4  \\
        & PolyLUT (NID Lite)\cite{andronicPolyLUTLearningPiecewise2023b} & DNN & 92\% & 3,336 & 686 & 0 & 0 & 529 & 9 & 3.00e+4  \\
        & PolyLUT (NID Lite)\tnote{1} & DNN & 92\% & 3,361 & 1,215 & 0 & 0 & 372 & 19 & 6.38e+4  \\
        & LogicNets\cite{umurogluLogicNetsCoDesignedNeural2020} & DNN & 91\% & 15,900 & — & 0 & 0 & 471 & 11 & 1.75e+5  \\
\hline
\end{tabular}

\begin{tablenotes}    
        \footnotesize               
        \item[1] These results were reproduced by the author using Xilinx FPGA \texttt{xcku060ffva1156-2-i}.               
\end{tablenotes}

\end{threeparttable} 
}

\end{center}
\end{table*}

\subsection{Datasets}

Table \ref{tabular} presents a comparison of the efficiency of various methods applied to different datasets (MNIST\cite{lidengMNISTDatabaseHandwritten2012}, JSC\cite{duarteFastInferenceDeep2018a}, and NID\cite{moustafaUNSWNB15ComprehensiveData2015}) on the implementations of DNNs and decision trees on FPGAs, including LogicNets\cite{umurogluLogicNetsCoDesignedNeural2020}, PolyLUT\cite{andronicPolyLUTLearningPiecewise2023b}, NeuraLUT\cite{andronicNeuraLUTHidingNeural2024}, PolyLUT-Add\cite{louPolyLUTAddFPGAbasedLUT}, DWN\cite{bacellarDifferentiableWeightlessNeural2024}, ReducedLUT\cite{cassidyReducedLUTTableDecomposition2024}, AmigoLUT\cite{wengGreaterSumIts2025}, and TreeLUT\cite{khataeiTreeLUTEfficientAlternative2025}. Here are the descriptions of the used datasets:

\begin{enumerate}
\item [1)] \textit{Handwritten Digit Recognition:}  
Advancements in autonomous driving, augmented reality, and edge computing have significantly increased the demand for real-time image classification. The ability to achieve low-latency inference enables a wide range of emerging applications in these rapidly evolving fields. The MNIST dataset, widely used as a benchmark in image classification, consists of handwritten digits represented as 28$\times$28 pixel grayscale images. These images are flattened into a 784-dimensional input vector, while the classification task involves mapping each input to one of 10 output categories, corresponding to digits 0 through 9.  

\item [2)] \textit{Jet Substructure Classification:}  
Efficient real-time inference and optimized resource utilization are critical for advancing high-energy physics research at the CERN Large Hadron Collider (LHC). Due to the high collision rate at the LHC, initial-stage data processing requires exceptional throughput to efficiently refine large volumes of sensor data. The jet substructure classification task involves analyzing 16 substructure properties to categorize jets into five distinct types. The JSC dataset is further divided into two categories based on accuracy: JSC-H (high accuracy, $>$74\%) and JSC-L (lower accuracy).  

\item [3)] \textit{Network Intrusion Detection:}  
Ultra-low-latency DNNs play a crucial role in cybersecurity, particularly in network intrusion detection systems (NIDS), which continuously monitor network traffic to identify and mitigate potential security threats. Given that fiber-optic networks can operate at speeds of up to 940 Mbps, NIDS solutions must support high-throughput processing while ensuring real-time threat detection. Additionally, privacy concerns necessitate on-chip FPGA-based implementations, which offer a secure and efficient alternative to cloud-based solutions. The NID dataset is designed for binary classification, where network packets are labeled as either safe (0) or malicious (1) based on 49 extracted features.          
\end{enumerate}

\subsection{Comparison of LUT-based DNNs and DTs}

The detailed model configurations for each method are provided in the Appendix. The performance of each approach across different datasets is analyzed as follows:

\textit{1) Handwritten digit recognition:} For the handwritten digit recognition task, compared to PolyLUT, the area-delay product was reduced by factors of 1.7$\times$, 7.6$\times$, 706.6$\times$, 1.4$\times$, 94.6$\times$, and 140.3$\times$ by NeuraLUT, PolyLUT-Add, DWN (sm), ReducedLUT, AmigoLUT-LogicNet-XS (2 models), and TreeLUT (II), respectively. 
DWN\cite{bacellarDifferentiableWeightlessNeural2024} achieved the highest accuracy of 98.3\%, the lowest LUT resource utilization, and the best area-delay product by training LUT-based models using gradient-based methods, which presents the improvements of fully leveraging the computational capabilities of LUTs. 
For the LUT compression method, ReducedLUT reduces the P-LUT utilization by up to 11\% compared to NeuraLUT. This smaller reduction stems from the fact that MNIST's L-LUT output bitwidth is limited to two, which restricts the ability of ReducedLUT to separate local variations into distinct sub-tables. 
TreeLUT\cite{khataeiTreeLUTEfficientAlternative2025} demonstrated the lowest latency among all evaluated models while maintaining high accuracy, making it particularly suitable for applications that require ultra-low latency without strict resource constraints. 

\textit{2) Jet substructure classification:} For the jet substructure tagging task, compared to PolyLUT (JSC-XL), the area-delay product was reduced by factors of 3.8$\times$, 8.5$\times$, 1911.5$\times$, 5.5$\times$, 12.1$\times$, and 5673.5$\times$ by NeuraLUT, PolyLUT-Add, DWN (md), ReducedLUT, AmigoLUT-NeuraLUT-S (32 models), and TreeLUT (II), respectively. Similarly, when compared to PolyLUT (JSC-M Lite), the area-delay product was reduced by factors of 4.4$\times$, 17.7$\times$, 4769.2$\times$, 4.7$\times$, 448.1$\times$, and 56.3$\times$ by NeuraLUT, PolyLUT-Add, DWN (sm), ReducedLUT, AmigoLUT-NeuraLUT-XS (4 models), and TreeLUT (II), respectively.
DWN \cite{bacellarDifferentiableWeightlessNeural2024} achieved the highest accuracy while utilizing the fewest LUTs for both JSC-L and JSC-H. Notably, DWN DWN attained 71.1\% accuracy with a latency of only 0.6 ns using just 20 LUTs, demonstrating exceptional efficiency. 
In terms of LUT compression techniques, ReducedLUT decreased P-LUT utilization by up to 36\% and 39\% for JSC-L and JSC-H when compared to NeuraLUT, with minimal impact on test accuracy.
TreeLUT exhibited the highest area-delay product while maintaining competitive classification accuracy and low latency with compact hardware utilization.  

\textit{3) Network intrusion detection:} For the network intrusion
task, compared to PolyLUT and PolyLUT-Add, TreeLUT (II) achieved reductions in the area-delay product by factors of 2.3$\times$ and 337.0$\times$, respectively. TreeLUT (I) achieved the highest accuracy of 93\% among all evaluated models, surpassing previous state-of-the-art approaches and demonstrating a powerful alternative to DNNs for inference acceleration on FPGAs

\section{Future Directions}

Despite significant advancements in LUT-based DNNs, many challenges remain in this field. The following are potential future research directions:

\begin{enumerate}

\item Development of Advanced Mathematical Tools for FPGA. Given FPGA's highly parallel computing capabilities and low power consumption, future research can explore mathematical tools specifically tailored for FPGA to enhance the computational efficiency and adaptability of bio-inspired DNNs.

\item Configurable and Learnable Activation Functions. Traditional DNN activation functions (e.g., ReLU, Sigmoid, Tanh) are typically fixed, whereas activation functions in biological DNNs dynamically change based on neurons and environmental conditions. Future research can focus on adaptive activation functions that allow FPGA hardware to dynamically adjust their shapes based on input distributions, thereby improving the generalization ability of DNNs.


\item Further Compression of LUT Resources. Existing LUT compression techniques, such as CompressedLUT and ReducedLUT, mainly rely on sub-table decomposition and don't-care term optimization but are still limited by the similarity within individual LUT patterns. Future research can explore inter-layer redundancies in DNNs, analyzing similarities across multiple LUTs to improve overall compression efficiency through logic merging or partial LUT sharing.

\item Neural Architecture Search (NAS) for LUT Optimization. Current LUT structure designs rely heavily on human expertise or predefined rules, making it difficult to achieve optimal configurations. Future research can leverage NAS (Neural Architecture Search) to automate LUT structure optimization and identify the most efficient LUT compositions. For example, FPGA-NAS\cite{abdelfattahCodesignNASAutomaticFPGA2020} has demonstrated automatic CNN structure search for FPGA, providing a reference for LUT-based DNN design automation.

\item Scalability. While existing LUT-based DNNs are primarily applied to small-scale networks, a key future direction is deploying \ac{LLM} on FPGA to expand its applications. The feasibility of FPGA-accelerated LLMs hinges on both model optimization and efficient hardware utilization. On the one hand, reducing LLM size through advanced optimization techniques, such as those proposed in DeepSeek R1\cite{deepseek-aiDeepSeekR1IncentivizingReasoning2025}, can make FPGA deployment more feasible. On the other hand, minimizing FPGA resource consumption requires exploring FPGA-specific quantization methods, such as the Microscaling (MX) data format\cite{rouhaniMicroscalingDataFormats2023}, which has been implemented and optimized by Ebby Samson et al.\cite{samsonExploringFPGADesigns2024}. Additionally, improving FPGA computational performance and designing architectures specifically tailored for LLM deployment are crucial. When a single FPGA cannot accommodate LLM requirements, further research is needed on resource allocation and communication strategies across multiple FPGAs to enable scalable and efficient deployment.

\end{enumerate}

\section{Summary}

The increasing demand for low-latency and energy-efficient DNN inference has driven the development of FPGA-based accelerators, which provide a balance between reconfigurability, power efficiency, and real-time performance. Traditional FPGA-based DNN implementations rely heavily on DSP blocks for multiply-accumulate operations, limiting scalability due to resource constraints. LUT-based DNNs have emerged as an alternative approach, leveraging FPGA LUTs to enhance computational efficiency and reduce inference latency. This survey examines the evolution of LUT-based DNN architectures, analyzing key design methodologies, trade-offs, and performance optimizations. Comparative evaluations across different application domains highlight both the benefits and limitations of these techniques. Future research directions are discussed, including advanced LUT compression strategies, scalable architectures, and the potential application of LUT-based methods in accelerating LLMs.

\bibliographystyle{IEEEtran}
\nocite{*}
\bibliography{LUT.bib}

\appendix
\clearpage

\begin{table*}[h]
\caption{Model setups of LogicNets}
\begin{center}
\begin{tabular}{c c c c c c}
\hline
\textbf{Dataset} & \textbf{Model Name}  & \textbf{Nodes per Layer}  & $\beta $ & $F$  & \textbf{Exceptions} \\
\hline
Jet substructure & JSC-L & 32, 64, 192, 192, 16 & 3 & 4 & $\beta_i$ = 4, $\beta_o$ = 7, $F_o$ = 5  \\
\hline
UNSW-NB 15 & NID-L & 593, 100, 100, 100 & 3 & 5 & $\beta_i$ = 2, $F_i$ = 7  \\
\hline

\end{tabular}
\label{tab0}
\end{center}
\end{table*}

\begin{table*}
\caption{Model setups of PolyLUT}
\begin{center}
\begin{tabular}{c c c c c c c}
\hline
\textbf{Dataset} & \textbf{Model Name}  & \textbf{Nodes per Layer}  & $\beta $ & $F$  & $D$ & \textbf{Exceptions} \\
\hline
MNIST & HDR & 256, 100, 100, 100, 100, 10 & 2 & 6 & 4 &   \\
\hline
Jet substructure & JSC-XL & 128, 64, 64, 64, 5 & 5 & 3 & 4 &   \\
\hline
Jet substructure & JSC-M Lite & 64, 32, 5 & 3 & 4 & 6 & $\beta_o$ = 7, $F_o$ = 2  \\
\hline
UNSW-NB15 & NID Lite & 686, 147, 98, 49, 1 & 2 & 7 & 4 & $\beta_o$ = 1  \\
\hline

\end{tabular}
\label{tab1}
\end{center}
\end{table*}

\begin{table*}
\caption{Model setups of PolyLUT-Add}
\begin{center}
\begin{tabular}{c c c c c c c}
\hline
\textbf{Dataset} & \textbf{Model Name}  & \textbf{Nodes per Layer}  & $\beta $ & $F$  & $D$ & \textbf{Exceptions} \\
\hline
MNIST & HDR-Add2 & 256, 100, 100, 100, 100, 10 & 2 & 4 & 3 &   \\
\hline
Jet substructure & JSC-XL-Add2 & 128, 64, 64, 64, 5 & 5 & 2 & 3 & $\beta_i$ = 7, $F_i$ = 1  \\
\hline
Jet substructure & JSC-M Lite-Add2 & 64, 32, 5 & 3 & 2 & 3 &   \\
\hline
UNSW-NB 15 & NID-Add2 & 100, 100, 50, 50, 1 & 2 & 3 & 1 & $\beta_i$ = 1, $F_i$ = 6, $\beta_o$ = 2, $F_o$ = 7  \\
\hline

\end{tabular}
\label{tab2}
\end{center}
\end{table*}

\begin{table*}
\caption{Model setups of NeuraLUT \& ReducedLUT}
\begin{center}
\begin{tabular}{c c c c c c c c c}
\hline
\textbf{Dataset} & \textbf{Model Name}  & \textbf{Nodes per Layer}  & $\beta $ & $F$  & $L$ & $N$ & $S$ & \textbf{Exceptions} \\
\hline
MNIST & HDR-5L & 256, 100, 100, 100, 10 & 2 & 6 & 4 & 16 & 2 &   \\
\hline
Jet substructure & JSC-2L & 32, 5 & 4 & 3 & 4 & 8 & 2 &   \\
\hline
Jet substructure & JSC-5L & 128, 128, 128, 64, 5 & 4 & 3 & 4 & 16 & 2 & $\beta_o$ = 7, $F_o$ = 2  \\
\hline

\end{tabular}
\label{tab1}
\end{center}
\end{table*}

\begin{table*}
\caption{Model setups of DWN}
\begin{center}
\begin{tabular}{c c c c}
\hline
\textbf{Dataset} & \textbf{Model Name}  & \textbf{Layers} & \textbf{Learning Rate}  \\
\hline
\multirow{3}{*}{MNIST} & DWN (sm; n=2) & 2x 6000 & 1e-2(30), 1e-3(30), 1e-4(30), 1e-5(10)   \\
& DWN (md; n=6) & 1000, 500 & 1e-2(30), 1e-3(30), 1e-4(30), 1e-5(10)   \\
& DWN (lg; n=6) & 2000, 1000 & 1e-2(30), 1e-3(30), 1e-4(30), 1e-5(10)   \\
\hline
\multirow{4}{*}{Jet substructure} & DWN (sm) & 1x 10 & 1e-2(14), 1e-3(14), 1e-4(4)   \\
& DWN (sm) & 1x 50 & 1e-2(14), 1e-3(14), 1e-4(4)   \\
& DWN (md) & 1x 360 & 1e-2(14), 1e-3(14), 1e-4(4)   \\
& DWN (lg) & 1x 2400 & 1e-2(14), 1e-3(14), 1e-4(4)   \\
\hline

\end{tabular}
\label{tab1}
\end{center}
\end{table*}

\begin{table*}
\caption{Model setups of AmigoLUT}
\begin{center}
\begin{tabular}{c c c c c c c}
\hline
\textbf{Dataset} & \textbf{Model Name}  & \textbf{Nodes per Layer}  & $\beta $ & $F$  & Layer bit width & Layer fan-in  \\
\hline
\multirow{3}{*}{MNIST} & AmigoLUT-LogicNet-XS & 1024, 1024, 128, 10 & 1 & 8 & 1, 1, 1, 4 & 8, 8, 8, 8  \\
& AmigoLUT-NeuraLUT (width = 16) & 600, 300, 300, 10 & 2 & 4 & 2, 2, 2, 2 & 4, 4, 4, 4 \\
& AmigoLUT-PolyLUT (degree = 3) & 600, 300, 300, 10 & 2 & 4 & 2, 2, 2, 2 & 4, 4, 4, 4  \\
\hline
\multirow{5}{*}{JSC} & AmigoLUT-LogicNet-S & 64, 32, 32, 32, 5 & 2 & 3 & 2, 2, 2, 2, 2 & 3, 3, 3, 3, 3  \\
& AmigoLUT-NeuraLUT-XS (width = 16) & 64, 5 & 2 & 3 & 2, 2 & 3, 3  \\
& AmigoLUT-NeuraLUT-S (width = 16) & 64, 5 & 4 & 3  & 2, 2 & 3, 3  \\
& AmigoLUT-PolyLUT-XXS (degree = 3) & 64, 5 & 2 & 3  & 2, 2 & 3, 3  \\
& AmigoLUT-PolyLUT-XS (degree = 3) & 64, 5 & 2 & 3  & 2, 2 & 3, 3  \\
\hline

\end{tabular}
\label{tab1}
\end{center}
\end{table*}

\begin{table*}[t]
\caption{Model setups of TreeLUT}
\begin{center}
\begin{tabular}{c c c c c c c c c}
\hline
\textbf{Dataset} & \textbf{Model Name}  & n\_estimators  & max\_depth  & eta & scale\_pos\_weight & w\_feature & w\_tree & Pipelining Parameters \\
\hline
\multirow{2}{*}{MNIST} & TreeLUT (I) & 30 & 5 & 0.8 & - & 4 & 3 & [0,1,1]\\
& TreeLUT (II) & 30 & 4 & 0.8 & - & 4 & 3 & [0,1,1]\\
\hline
\multirow{2}{*}{JSC} & TreeLUT (I) & 13 & 5 & 0.8 & - & 8 & 4 & [0,1,1]\\
& TreeLUT (II) & 10 & 5 & 0.3 & - & 8 & 2 & [0,1,0]\\
\hline
\multirow{2}{*}{NID} & TreeLUT (I) & 40 & 3 & 0.6 & 0.3 & 1 & 5 & [0,0,1]\\
& TreeLUT (II) & 10 & 3 & 0.8 & 0.2 & 1 & 5 & [0,0,1]\\
\hline

\end{tabular}
\label{tab1}
\end{center}
\end{table*}

\end{document}